\documentclass[10pt, twocolumn, aps, superscriptaddress, prl ]{revtex4-1}

\DeclareUnicodeCharacter{202F}{~}

\usepackage[colorlinks=true,urlcolor=blue,citecolor=blue,linkcolor=red]{hyperref}
\usepackage{graphicx}
\usepackage{physics}
\usepackage{xcolor}
\usepackage{wrapfig}
\usepackage{booktabs}
\usepackage{amsmath}
\usepackage{amssymb}
\usepackage{physics}
\usepackage{braket}
\usepackage[dvipsnames]{xcolor}
\usepackage[normalem]{ulem}
\usepackage{hyperref}
\usepackage{bbold}

\usepackage[normalem]{ulem}
\usepackage{xcolor}

\begin{document}
	\title{Effect of Energy Extensivity on the Performance of 
		Open Quantum Interferometers}

\author{\v{Z}an Kokalj}\email{zan.kokalj@phd.units.it}
\affiliation{University of Trieste, Strada Costiera 11, I-34151 Trieste, Italy}
\author{Tommaso Favalli}\email{tommaso.favalli@units.it}
\affiliation{University of Trieste, Strada Costiera 11, I-34151 Trieste, Italy}
\author{Andrea Trombettoni}\email{atrombettoni@units.it}
\affiliation{University of Trieste, Strada Costiera 11, I-34151 Trieste, Italy}

\begin{abstract}
Studying the performance of a quantum interferometer coupled to an external environment is a problem of conceptual and practical importance. If we consider a quantum interferometer featuring Heisenberg-limited sensitivity, then a typical result is that introducing coupling with the environment degrades the sensitivity to the shot-noise limit. Here we argue that this result crucially depends on whether the interferometer-environment coupling term is subject (or not) to the so-called Kac rescaling that restores extensivity, i.e., whether the coupling Hamiltonian is extensive or not.
We present results of the Lindblad equation in the presence and absence of Kac rescaling of the coupling constant.
Our results show that for a linear coupling and a harmonic model of the environment, often used in modeling of a quantum interferometer coupled with an environment, the Heisenberg-limited sensitivity may be restored after the Kac rescaling. This result points out the need and the importance to characterize the (model of the) environment of the interferometer at hand.
\end{abstract}
\maketitle

\onecolumngrid

{\it Introduction:} When considering a physical system in its thermodynamic limit, there are certain properties one expects the system to satisfy. Among them, one of paramount importance is the extensivity of the energy \cite{callen1980thermodynamics,huang1987statistical,ruelle1999statistical}. If it is not present, consequences may include non-additivity of the entropy, non-equivalence of thermodynamic ensembles, negativity of response functions such as the specific heat, and non-convexity of the space of accessible thermodynamic parameters \cite{campa2009statistical}. Despite these features, the study of non-extensive systems is an active field of research, especially in systems with long-range interactions \cite{campa2009statistical,campa2014physics} such as gravitational \cite{he2012equilibrium} and quantum atomic, molecular, and optical systems \cite{defenu2023longrange}. 

A typical instance of broken energy extensivity is provided by the Ising model with an all-to-all interaction \cite{parisi1988statistical,nishimori2011elements}. A way to restore the extensivity is to perform the so-called Kac rescaling \cite{kac1963van}, where the coupling strength is rescaled by the number of particles $N$ in order to make the ground state energy scale linearly with the number of spins or particles. One can similarly proceed if the interaction is not all-to-all, but instead decaying as $1/r^{\alpha}$, with $r$ being the distance between constituents of the system (particles or spins). If $d$ is the spatial dimension of the system,  the system's energy is extensive for $\alpha>d$, while for $\alpha \le d$ the interaction term has to be divided by the Kac prefactor to ensure that the ground state energy per particle or spin is finite. Several studies in the literature characterize the effects of imposing (or relaxing) the requirement of energy extensivity in both classical and quantum systems \cite{mori2010analysis,mori2012equilibrium,botzung2021effects,granet2023exact}, see the recent review \cite{kastner2025long}. 

An important point to be noticed is that, while the energy extensivity is a physical property, the Kac rescaling is a mathematical operation we perform on the Hamiltonian of the system ${\cal S}$ under consideration to impose energy extensivity in the thermodynamic limit $N \to \infty$, where $N$ is the number of particles of the system. There may well be systems -- such as trapped ions interacting as $\sim 1/r^\alpha$ with $\alpha < d$ \cite{monroe2021programmable} -- where the Kac prefactor is simply not present in the Hamiltonian. As soon as $N$ is finite and the (effective) Hamiltonian $H$ of the system is fixed, we can use it without Kac rescaling. From a more formal standpoint, not performing the Kac rescaling implicitly assumes that the system ${\cal S}$ at hand is in fact a subsystem of a larger ${\cal S}_{tot}$. 
If $N_{tot}$ denotes the number of particles of the overall system ${\cal S}_{tot}$, then for large $N_{tot}$ the usual extensivity of energy and entropy will hold, as discussed in textbooks \cite{huang1987statistical}, allowing us to apply the standard framework of statistical mechanics. So, if $N_{tot}$ is large and the Hamiltonian $H_{tot}$ is not extensive, we have to restore extensivity of the overall system, but not (necessarily) for the subsystem ${\cal S}$ at a fixed $N$. For the sake of simplicity, we will refer to ${\cal S}$ as the system, even though it is a subsystem of an ${\cal S}_{tot}$. 
In this paper we are interested in the case where the system ${\cal S}$ is a quantum interferometer coupled with an environment ${\cal E}$, so that ${\cal S}_{tot}$ consists of ${\cal S}$ and ${\cal E}$. 

{\it Main idea:} The reason for such an interest is based on the substantial effort in the field of quantum sensing to study quantum interferometers and characterize the advantage one can achieve using entanglement \cite{degen2017quantum}, a major example being provided by metrology with nonclassical states of cold atomic ensembles \cite{pezze2018quantum}. 
As is well known \cite{pezze2018quantum}, the sensitivity scales as $1/\sqrt{N}$ in the shot-noise limit (i.e., when the entanglement is absent or not useful as a quantum resource), while in the quantum Heisenberg limit the sensitivity scales as $1/N$. Given the importance, both conceptual and practical, of getting to the Heisenberg limit \cite{giovannetti2004quantum,pezze2018quantum}, and the difficulty to reach it \cite{demkowiczdobrzański2012elusive}, the study of the performance of quantum interferometers (such as ultra-cold ones), has attracted considerable attention, see the reviews \cite{pezze2018quantum,wolswijk2025trapping}.
Since sensing devices are subject to noise, decoherence, and losses, limiting their performance, a significant body of research has focused on the characterization of the degradation of sensitivity of quantum interferometers \cite{demkowiczdobrzanski2009quantum,escher2011general,escher2012quantum,das2012universal,demkowiczdobrzański2012elusive,demkowiczdobrzanski2014using,yue2014quantum,kolodynski2014precision,demkowiczdobrzanski2017adaptive,albarelli2018restoring,albarelli2022quantum,gorecki2022quantum,bao2022multichannel,gorecki2025interplay,das2025quantum,kurdzialek2025universal}. A general framework for estimating the precision limit of noisy quantum interferometers has been developed \cite{escher2011general}, and a considerable amount of effort focused on the development of techniques to mitigate and contrast effects of noise and losses \cite{demkowiczdobrzanski2014using,albarelli2018restoring,albarelli2022quantum}. Our goal here is to discuss how the extensivity of the Hamiltonian describing the coupling of a quantum interferometer to the environment may affect its performance. Since here we do not enter into the detailed description of losses and noise  \cite{demkowiczdobrzanski2009quantum,demkowiczdobrzanski2017adaptive,yue2014quantum,  gorecki2022quantum,kurdzialek2025universal}, but instead consider the  interferometers described by open system dynamics \cite{breuer2002theory}, we prefer to use the term open quantum interferometry, rather than noisy or lossy quantum 
interferometry.

In open quantum systems, a typical approach is to consider the total system ${\cal S}_{tot}$, consisting in our case of the interferometer ${\cal S}$ and an environment ${\cal E}$, typically modeled as a collection of harmonic oscillators.
In such a context, maintaining the extensivity of the ground state energy of the total system may demand an appropriate rescaling of the parameters of the global Hamiltonian. When the environment is represented by harmonic oscillators, energy extensivity may be restored by rescaling the coupling strength between the system and the environment.

Nonetheless, as discussed above, we warn that rescaling may not always be appropriate, particularly when the focus is on investigating specific properties of the subsystem. In some cases, the rescaling procedure may obscure or eliminate the very phenomena of interest. For instance, in the Caldeira–Leggett model \cite{caldeira1981influence,leggett1987dynamics} where a harmonic oscillator is coupled to a bath of harmonic oscillators, if the primary interest lies in the damping of the oscillator induced by the environment, then rescaling the coupling may suppress this effect and then the issue of the extensivity of the coupling is typically not discussed.

For the case of an interferometer, since one has $N \ll N_{tot}$, if the number of particles $N$ of the interferometer ${\cal S}$ is fixed, then it is certainly meaningful not to restore energy extensivity. However, when one wants to study the scaling of the sensitivity when $N$ in turn is increasing, then one may ask what are the effects of restoring the energy extensivity of the total Hamiltonian and of the coupling term between ${\cal S}$ and ${\cal E}$. 

The sensitivity of quantum interferometers coupled with an environment can be obtained by studying the Lindblad equation, where the environment has been traced out \cite{breuer2002theory}, 
and it reveals that , for large $N$, the Heisenberg limit is typically lost, and the sensitivity returns to the shot-noise scaling, see \cite{escher2011general} and as well below. If one has a Heisenberg-limited closed quantum interferometer (not coupled to an environment)  and then adds a coupling term between ${\cal S}$ and ${\cal E}$ with the coefficient $g$, a typical result is that the value of $N$ at which the Heisenberg scaling is no longer observed, increases with decreasing $g$. The question we want to address is therefore, if and how the Kac rescaling of the coupling ${\cal S}$ - ${\cal E}$ alters this typical result.
We will then focus on analyzing the performance of an open quantum interferometer under the influence of noise arising from its interaction with an external environment. To fix the notations and the setup, the system of interest is the quantum interferometer itself, having $N$ bosonic particles, while the environment is modeled as a generic bath comprised of $M$ harmonic oscillators.

{\it The model:}
A common way to model a quantum interferometer is given by the so-called linear two-mode (TM) model \cite{pezze2018quantum}:
\begin{equation}
H_{TM}= -J(a^{\dagger} b + b^{\dagger} a) + \frac{\delta}{2} (a^{\dagger} a - b^{\dagger} b),  
\label{TM}
\end{equation}
where the modes $a$ and $b$ refer to annihilation of particles in the two wells \cite{javanainen1996quantum}, and the coefficients 
$J$ and $\delta$ are the tunneling amplitude and the energy gap between the wells, respectively. 

By tuning $J$ and $\delta$ in time, the model (\ref{TM}) can be used to realize an atom interferometer. A typical interferometric protocol consists of: {\it 1)} preparing the system in an initial state;
{\it  2)} lowering the barrier between the wells (increasing $J$) with 
$\delta=0$ for a duration of $T_{BS}$ (beam-splitting); {\it 3)} 
turning off the tunneling, $J=0$ (i.e., decoupling the wells), and setting $\delta\neq0$ for the duration of holding time $T_H$ (phase accumulation stage); {\it 4)} performing the beam-splitter again; {\it 5)} measuring an operator $O$ after the second beam-splitting. By repeating the measurement process many times, one can estimate the parameter $\delta$ and determine its uncertainty $\Delta \delta$, often referred to as the sensitivity of the measurement. 

As an example, the initial state could be $|N,0\rangle$ (i.e., all the particles in the mode $a$), $|N/2,N/2\rangle$ (the Twin Fock (TF) state) or $\frac{1}{\sqrt{2}}\left(|N,0\rangle+|0,N\rangle \right)$, the NOON state. For $|N,0\rangle$, one can measure the operator 
$a^\dagger a-b^\dagger b$, i.e., the difference in the number of atoms in the two modes; while for the NOON and TF initial states one can measure the parity operator $\Pi_b = e^{i\pi n_b}$, where $n_b=  b^\dagger b$.

Since the early development of elementary quantum interferometric models, there has been a significant interest in understanding how this sensitivity scales with the number of particles (or probes \cite{escher2011general}) involved. In particular, the focus has been on how quantum resources can yield a quantum enhancement in sensitivity, surpassing the classical shot-noise limit, which scales as 
$1/\sqrt N $ \cite{caves1981quantum}. It has been shown that, for a closed quantum system, the sensitivity can reach the so-called Heisenberg limit, where $\Delta \delta \propto 1/N$ \cite{caves1981quantum}. A necessary (though not sufficient) condition for achieving this scaling is that the initial state exhibits quantum entanglement \cite{pezze2009entanglement}. 
The sensitivity $\Delta \delta$ is lower-bounded by the Cram\'{e}r-Rao relation as $\Delta \delta\geq 1/\sqrt{F_Q}$, where $F_Q$ is the quantum Fisher Information \cite{braunstein1994statistical}, which can be calculated from the final (after the second beam-splitting) density matrix of the system. 
The calculation of $F_Q$ has been extended to open quantum systems: it can be done by knowing the Kraus operators and applying the formalism described in \cite{escher2011general}. Details on the calculation of $F_Q$ are given in Appendix~A.

{\it Scaling analysis:} Let us now present our main results for the study of the energy extensivity of the two mode model coupled to an environment. It is instructive to first examine the simpler case of a single harmonic oscillator coupled to an environment. We start with the simplest case of an environment consisting of just one harmonic oscillator, with the Hamiltonian reading
\begin{equation}
H_{tot} = H_S + H_E + H_I = \omega_a a^{\dagger} a + \omega_R R^{\dagger} R + g(a^{\dagger} R + R^{\dagger} a ), 
\end{equation}
where $H_S$, $H_E$ and $H_I$ are the system, the single-oscillator environment and the interaction Hamiltonian respectively; $g$ is the coupling strength and $\omega_{a}, \omega_{R}$ the system and environment harmonic oscillator frequencies, respectively. Furthermore, we fix the average number of particles in the ground state $|GS \rangle$ of the total system to $\langle a^{\dagger} a\rangle = N$, and in the environment to $\langle R^{\dagger} R\rangle = M$. If we now calculate the expectation value of $H_I$ with respect to the $|GS \rangle$ (see Appendix B), we get:
\begin{equation}\langle H_I \rangle = \langle GS| 
H_I |GS \rangle =
2g \sqrt{N} \sqrt{M}.\end{equation}
We continue by extending the environment from $1$ to $M$ harmonic oscillators, i.e., changing the environment Hamiltonian to \( \sum_{i=1}^{M} \omega_i R_i^{\dagger} R_i\) and the interaction Hamiltonian to \(H_I=\sum_{i=1}^{M} g_i (R_i^{\dagger} a + h.c.)\), where $M$ is now the number of environment harmonic oscillators and $g_i$ is the coupling strength between the system and the $i$-th environment harmonic oscillator. Now, we fix particle number in each environment harmonic oscillator to $\langle R_i^{\dagger} R_i\rangle = M_i$. Doing the same calculation for the interaction energy ground state (see Appendix C), assuming $M_i = M_{AVG}$ and $g_i=g$  $\forall i$, and (crucially) posing
\begin{equation}
N = \epsilon M
\label{epsilon}
\end{equation}
where $\epsilon \ll 1$, we get:
\begin{equation}\langle H_I \rangle = 
\frac{2g}{\epsilon} 
N^{\frac{3}{2}} \sqrt{M_{AVG}} \, .
\label{eq5}
\end{equation}
We notice that the interaction energy scales as $N^{\frac{3}{2}}$ with $N$. Therefore, {\it if $\epsilon$ is finite}, in order to restore extensivity, $g$ has to be divided by $\sqrt{N}$. Finally, we observe that if we generalize the sum to an integral by considering \(H_I=\int_{0}^{\infty} d\omega \rho(\omega)g(\omega) (R_\omega^{\dagger} a  + h.c.)\) with $\rho$ density of states and $R_\omega$ the annihilation operator related to the oscilator mode $\omega$, then one gets again (\ref{eq5}) with $g(\omega)=g$, $M_{\omega}=M_{AVG}$ $\forall \omega$, and, as before, $\int_{0}^{\infty} d\omega \rho(\omega) = M = N/\epsilon$.

We now consider the TM model (\ref{TM}), focusing on $J=0$ and $\delta \neq 0$, i.e., during the phase accumulation stage of the interferometer. For the sake of illustration (and as representative), we introduce a simple form of the coupling where the two modes may exchange particles with the environment modes:
\begin{equation}
H_{tot} = H_S + \sum_{i=1}^{M} \omega_i R_i^{\dagger} R_i+ \sum_{i=1}^{M} g_i \left[(R_i^{\dagger} a +R_i^{\dagger} b) + h.c.\right], 
\end{equation}
where $H_S=H_{TM}(J=0)=\frac{\delta}{2} (a^\dagger a - b^\dagger b)$.
Again taking 
$ M_i = M_{AVG}$, $g_i=g$,
$N = \epsilon M$ according to \ref{epsilon}), and assuming equal population of both wells, one gets:
\begin{equation} 
\langle H_I \rangle = 4 \sqrt{N/2} \sum_{i=1}^{M} g_i \sqrt{M_i}=  \frac{2\sqrt{2} \,g}{\epsilon}  
N^{\frac{3}{2}} \sqrt{M_{AVG}}.
\label{eq10}
\end{equation}
It can be argued then that $g$ should be divided by $\sqrt{N}$, in order to restore the extensivity. As discussed for the single mode coupled to the environment, passing to a continuous bath gives the same result for $\langle H_I \rangle$ as (\ref{eq10}), see Appendix D.

Before dealing with the open quantum interferometer, some comments are in order: {\it 1)} The rescaling of $g$ is required to make $H_I$ extensive, which makes the total Hamiltonian $H_{tot}$ extensive as well. Notice that, 
since $\langle H_I \rangle \propto \sqrt{N} M$, with finite 
$\epsilon$ the rescaling amount needed to make $H_I$ extensive with respect to the total number of particles of the system, which is $N+M$, is $\sqrt N$. {\it 2)} As discussed in the introduction, if $N$ is finite and fixed (i.e., we are not studying the scaling of sensitivity with $N$), $g$ need not be rescaled. {\it 3)} Finally, it could be the case that both $N$ and $M$ diverge, but their ratio $N/M \to 0$ for $N,M \to \infty$. This occurs when $N$ scales as $M^\sigma$, with $\sigma <1$. In that case one has to rescale $g$ by $N^{1/2+1/\sigma-1}$.

{\it Open quantum interferometer:} So far we have only considered the ground state of our Hamiltonian describing the quantum interferometer during the holding phase, coupled to an environment. We now study the evolution of the interferometer during the phase accumulation stage and examine how the final sensitivity measured depends on the scaling of the coupling strength. To this end, we consider the continuous version of (\ref{eq10}):
\begin{equation}
H_{tot} = H_S+H_E+\int d \omega \rho(\omega) g(\omega) \left[\left(R^{\dagger}_\omega a +R_\omega^{\dagger} b\right) + h.c.\right],
\end{equation}
where $H_I=\int d \omega \rho(\omega) \omega R^{\dagger}_\omega R_\omega$. We write $H_S=H_{TM}(J=0)+V_0 (a^\dagger a+b^\dagger b)$, writing explicitly the chemical potential. Following \cite{louisell1967solutions} and \cite{schlosshauer2019quantum}, by tracing out the environment and  assuming the Born-Markov approximation \cite{louisell1967solutions,breuer2002theory,schlosshauer2019quantum}, we obtain the following Lindblad equation:
\begin{equation} \frac{\partial\rho}{\partial t}= -i[ H_S, \rho]+\mathcal{D}[\rho], \end{equation}
where $\mathcal{D}[\rho] = \gamma ( L \rho  L ^{\dagger}-\frac{1}{2} \{ L ^{\dagger} L , \rho\})$ with $L = \frac{1}{\sqrt 2}(a+b) \equiv \alpha$ and $\gamma \approx 4 \pi g^2 \rho \Big|_{\omega=V_0}$, where we have assumed the low temperature limit (see \cite{louisell1967solutions} and Appendix E). Since $\gamma \propto g^2$, one has to rescale $\gamma$ by $N$ as a consequence of the restoration of energy extensivity (if we decide to restore it). We will use this approximation in the following to produce numerical results.

{\it Results:} The Lindblad equation obtained above describes the reduced dynamics of the open quantum system. For low values of $N$ one can obtain the dynamics of $\rho(t)$ analytically via vectorization \cite{qvarfort2025solving}. 
Since we are ultimately interested in the scaling of sensitivity with $N$, we resort to approximate numerical methods to calculate the dynamics for large $N$ \cite{breuer2002theory,schaller2014open}. We simulate the Fock space up to a certain cut-off $N_{max}$ and for the initial state we choose $|N,0\rangle$, $|N/2,N/2\rangle$ (the TF state) or $\frac{1}{\sqrt{2}}\left(|N,0\rangle+|0,N\rangle \right)$ (the NOON state). Since, for our Lindblad operator, the dissipator only induces particle loss, we set the cutoff to $N_{max}=N$. 
Instead of calculating the sensitivity $\Delta \delta$ of the measured parameter $\delta$, we calculate its Cram\'{e}r-Rao lower bound (CRLB) \cite{pezze2009entanglement}, because it does not depend on the choice of POVM measurement.

\begin{figure}[b]
\centering
\includegraphics[width=0.8\textwidth]{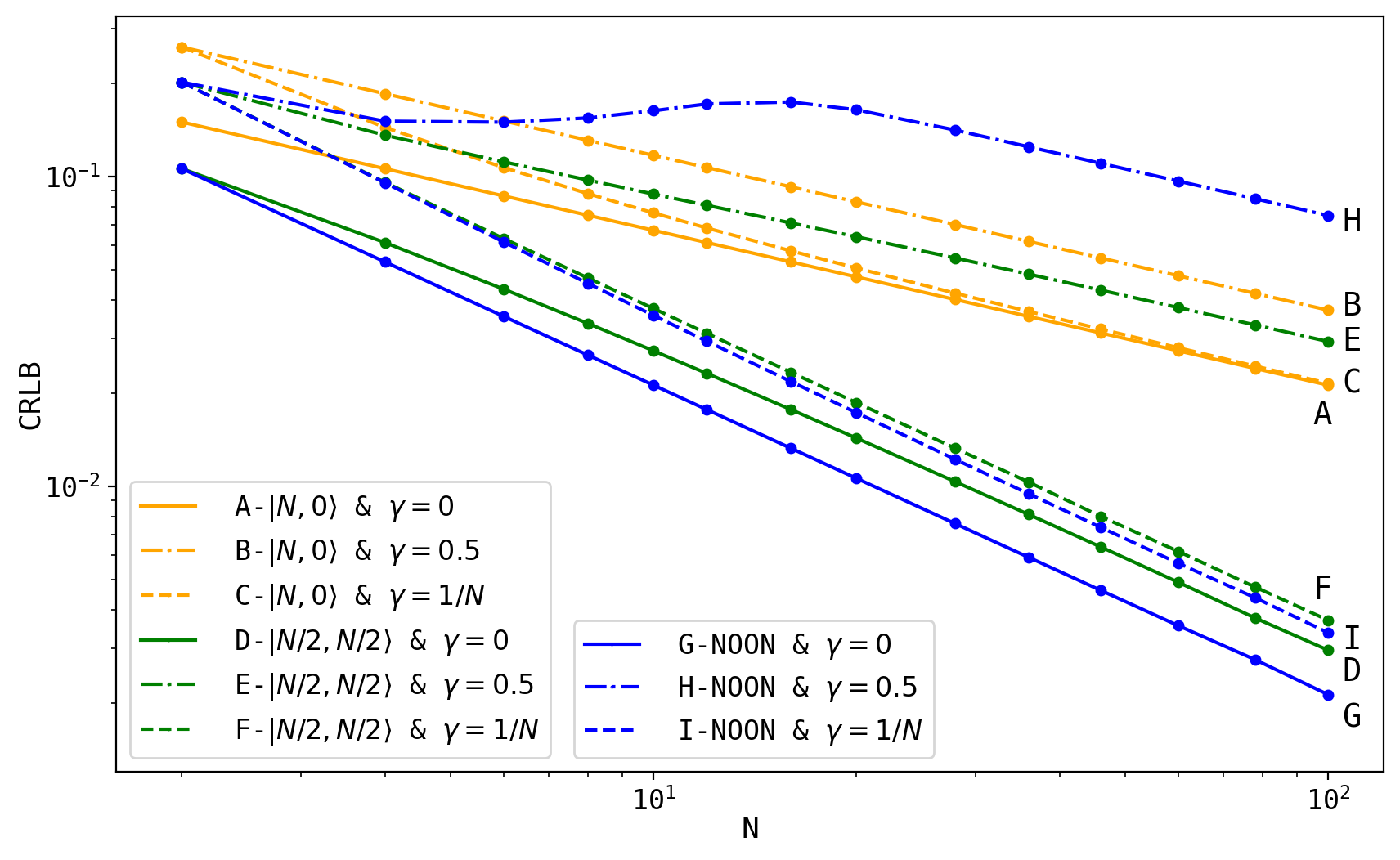} 
\caption{Effect of rescaling $\gamma$. Figure shows the CRLB of sensitivity $\Delta \delta$  and its scaling with the number of particles $N$ for an open quantum interferometer governed by the Lindblad equantion with a particle conserving Lindblad operator $L=(a+ b)/\sqrt{2}$. Three different initial states are considered in the simulation, and the solid, dash-dotted, dotted lines correspond to $\gamma_1=0$, $\gamma_2=0.5$, $\gamma_3 N=1$ respectively. We see that in the NOON and TF cases scaling behavior is restored to Heisenberg-limited scaling by rescaling of the noise parameter.}
\label{fig:Figure1}
\end{figure}

For each initial state we simulate the dynamics and calculate the CRLB of the process for three cases:  $\gamma_1 = 0$ (noiseless case), $\gamma_2 = 0.5$ (noise parameter not rescaled with N) and $\gamma_3 = 1/N$ (noise parameter rescaled with $N$, as discussed above). Since the TF state is only defined for even $N$, the first considered $N$ is $2$. In order that the CRLB curves as a function of $N$ in the second and third case start from the same value, we decided to choose $\gamma_3=2\gamma_2/N=1/N$ (however, we would reach the same conclusions also for $\gamma_3=\gamma_2/N=0.5/N$). 

As can be seen from the solid lines on the Fig.~\ref{fig:Figure1} the noiseless case $\gamma_1=0$ follows the usual $1/\sqrt{N}$ shot-noise scaling of CRLB for the $|N,0\rangle$ initial state, and the Heisenberg-limited $1/{N}$ scaling of CRLB for the TF and NOON states, as expected \cite{pezze2018quantum}.

Dash-dotted lines on Fig.~\ref{fig:Figure1} show the behaviour of CRLB when the noise parameter is fixed at $\gamma_2 = 0.5$. Here we observe that the scaling for all initial states changes to the shot-noise scaling $1/\sqrt{N}$. This was to be expected from previous discussions on the noisy quantum interferometry \cite{escher2011general}.
Dashed lines on the Fig.~\ref{fig:Figure1} show the same simulation, but taking into account the rescaling of $\gamma$, i.e., $\gamma=\gamma_3$. We see that we restore the Heisenberg-limited scaling in TF and NOON case, and shot-noise scaling in the $|N,0\rangle$ case. The recovery of the Heisenberg-limited scaling is also consistent with the recent studies on optimal signal-to-noise ratio in the presence of noise \cite{gorecki2025interplay}, where is shown that $F_Q \propto N/\gamma$ for frequency-estimation protocols.

\begin{figure}[t]
\centering
\includegraphics[width=0.8\textwidth]{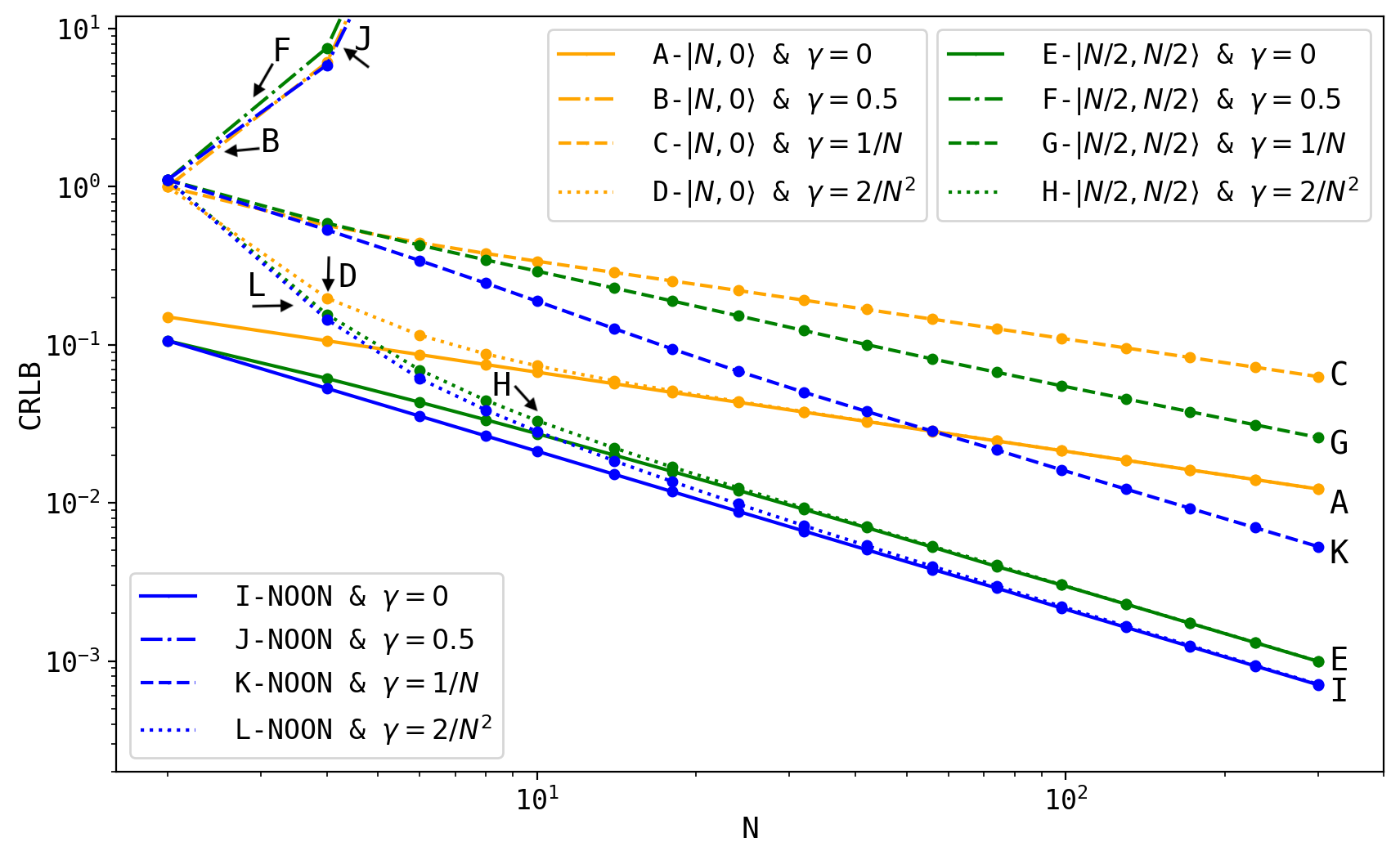} 
\caption{CRLB of the sensitivity $\Delta \delta$  and its scaling with the number of particles $N$  is plotted for an open quantum system governed by the Lindblad equation with a particle conserving Lindblad operator $L=b^\dagger a$. We consider three different initial states in the simulation, and the solid, dash-dotted, dashed and dotted  lines correspond to $\gamma_1=0$, $\gamma_2 N=1$, $\gamma_3 N^2=2$ and $\gamma_4=0.5$, respectively. We show that in the NOON and $|N,0\rangle$ cases their scaling behavior is restored by rescaling with $N$ of the noise parameter, whereas for the TF state, rescaling with $N$ results in an improved sensitivity although the Heisenberg-limited scaling is not restored. When rescaling with $N^2$, the $\gamma=0$ behavior is always restored.}
\label{fig:Figure2}
\end{figure}

In Fig.~\ref{fig:Figure2} we present the simulations of the Lindbladian for the TM model, but now with the Lindblad operator $L=b^\dagger a$, which conserves the system particle number. All the other parameters are the same as on Fig.~\ref{fig:Figure1}, except we add plots of evolution produced by rescaling with $N^2$ as well (dotted lines). This is because the ground state interaction energy of the microscopic Hamiltonian corresponding to $L=J_-$ scales as $\langle H_I \rangle \propto N^2 g$, which requires $g$ to be scaled by $N$, or $\gamma$ by $N^2$. We notice that CRLB at $\gamma=0.5$  immediately starts increasing with $N$ and the Heisenberg-limited scaling is only restored for the NOON initial state after rescaling with $N$ and for both TF and NOON initial states, if the rescaling is done by $N^2$. The simulation with TF as the initial state, on the other hand, leads to a CRLB scaling that is between shot-noise and Heisenberg-limited scaling after $\gamma$ is rescaled by just $N$. Therefore, we conclude that the $1/N$ rescaling {\it may} restore the $1/N$ Heisenberg scaling, but not necessarily.

{\it Conclusions:} Given the importance of studying 
the performance of a quantum interferometer coupled with an external environment, we investigated the effect of imposing the restoration of the energy extensivity of the Hamiltonian coupling the interferometer and the environment on its performance.
Without the rescaling, a typical result is that the sensitivity is degraded to the shot-noise limit by the coupling with the environment. We argued that this result crucially depends on whether the interferometer-environment coupling term is (or, it has to be) subject, or not, to the Kac rescaling to restore energy extensivity, i.e., whether the coupling Hamiltonian $H_I$ is extensive or not.
We presented results of the Lindblad equation 
for the same microscopic model with and without the Kac rescaling of the coupling constant. We found that the rescaling may restore the Heisenberg-limited sensitivity. 

We clarified that if $N$ is fixed, the Kac rescaling does not have to be done. Since it is the total energy of the entire system that includes the interferometer that has to be extensive, our conclusion is not that the rescaling has to always be done and that the Heisenberg limit is always restored, but rather that: {\it 1)} In specific cases, it is the accurate characterization of the environment guiding the choice whether to restore the energy extensivity of $H_I$; and {\it 2)} Not imposing the energy extensivity on the coupling Hamiltonian may overestimate the degradation of the sensitivity determined by the environment. In this direction, both theoretical and experimental studies of quantum interferometers coupled to a tunable environment (e.g., with a realistic form of the dependence of the coupling $g$ on $\omega$) would be very interesting in the light of the findings of the present paper.

\vspace{0.2cm}
\textit{Acknowledgements:} Discussions with F. Albarelli, F. Benatti, L. Pezz\'{e} and A. Smerzi are gratefully acknowledged. \v{Z}.K. and A.T. acknowledge funding from the European Union’s Horizon Europe Research and Innovation Programme under the Marie Sk\l{}odowska-Curie Doctoral Network MAWI (Matter-Wave Interferometers) 
under the grant agreement No. 101073088. T.F. and A.T. thank the Project \lq\lq National Quantum Science and Technology Institute – NQSTI\rq\rq \, Spoke 3: \lq\lq\ Atomic, molecular platform for quantum technologies\rq\rq.

\bibliography{references}

\begin{thebibliography}{47}%
\makeatletter
\providecommand \@ifxundefined [1]{%
 \@ifx{#1\undefined}
}%
\providecommand \@ifnum [1]{%
 \ifnum #1\expandafter \@firstoftwo
 \else \expandafter \@secondoftwo
 \fi
}%
\providecommand \@ifx [1]{%
 \ifx #1\expandafter \@firstoftwo
 \else \expandafter \@secondoftwo
 \fi
}%
\providecommand \natexlab [1]{#1}%
\providecommand \enquote  [1]{``#1''}%
\providecommand \bibnamefont  [1]{#1}%
\providecommand \bibfnamefont [1]{#1}%
\providecommand \citenamefont [1]{#1}%
\providecommand \href@noop [0]{\@secondoftwo}%
\providecommand \href [0]{\begingroup \@sanitize@url \@href}%
\providecommand \@href[1]{\@@startlink{#1}\@@href}%
\providecommand \@@href[1]{\endgroup#1\@@endlink}%
\providecommand \@sanitize@url [0]{\catcode `\\12\catcode `\$12\catcode `\&12\catcode `\#12\catcode `\^12\catcode `\_12\catcode `\%12\relax}%
\providecommand \@@startlink[1]{}%
\providecommand \@@endlink[0]{}%
\providecommand \url  [0]{\begingroup\@sanitize@url \@url }%
\providecommand \@url [1]{\endgroup\@href {#1}{\urlprefix }}%
\providecommand \urlprefix  [0]{URL }%
\providecommand \Eprint [0]{\href }%
\providecommand \doibase [0]{http://dx.doi.org/}%
\providecommand \selectlanguage [0]{\@gobble}%
\providecommand \bibinfo  [0]{\@secondoftwo}%
\providecommand \bibfield  [0]{\@secondoftwo}%
\providecommand \translation [1]{[#1]}%
\providecommand \BibitemOpen [0]{}%
\providecommand \bibitemStop [0]{}%
\providecommand \bibitemNoStop [0]{.\EOS\space}%
\providecommand \EOS [0]{\spacefactor3000\relax}%
\providecommand \BibitemShut  [1]{\csname bibitem#1\endcsname}%
\let\auto@bib@innerbib\@empty
\bibitem [{\citenamefont {Callen}(1980)}]{callen1980thermodynamics}%
  \BibitemOpen
  \bibfield  {author} {\bibinfo {author} {\bibfnamefont {H.~B.}\ \bibnamefont {Callen}},\ }\href@noop {} {\bibfield  {journal} {\bibinfo  {journal} {John Wiley\& Sons}\ }\textbf {\bibinfo {volume} {2}} (\bibinfo {year} {1980})}\BibitemShut {NoStop}%
\bibitem [{\citenamefont {Huang}(1987)}]{huang1987statistical}%
  \BibitemOpen
  \bibfield  {author} {\bibinfo {author} {\bibfnamefont {K.}~\bibnamefont {Huang}},\ }\href@noop {} {\emph {\bibinfo {title} {Statistical Mechanics}}},\ \bibinfo {edition} {2nd}\ ed.\ (\bibinfo  {publisher} {John Wiley \& Sons},\ \bibinfo {address} {New York},\ \bibinfo {year} {1987})\BibitemShut {NoStop}%
\bibitem [{\citenamefont {Ruelle}(1999)}]{ruelle1999statistical}%
  \BibitemOpen
  \bibfield  {author} {\bibinfo {author} {\bibfnamefont {D.}~\bibnamefont {Ruelle}},\ }\href@noop {} {\emph {\bibinfo {title} {Statistical Mechanics: Rigorous Results}}}\ (\bibinfo  {publisher} {World Scientific},\ \bibinfo {address} {Singapore},\ \bibinfo {year} {1999})\ \bibinfo {note} {reprint of original 1969 edition}\BibitemShut {NoStop}%
\bibitem [{\citenamefont {Campa}\ \emph {et~al.}(2009)\citenamefont {Campa}, \citenamefont {Dauxois},\ and\ \citenamefont {Ruffo}}]{campa2009statistical}%
  \BibitemOpen
  \bibfield  {author} {\bibinfo {author} {\bibfnamefont {A.}~\bibnamefont {Campa}}, \bibinfo {author} {\bibfnamefont {T.}~\bibnamefont {Dauxois}}, \ and\ \bibinfo {author} {\bibfnamefont {S.}~\bibnamefont {Ruffo}},\ }\href {\doibase 10.1016/j.physrep.2009.07.001} {\bibfield  {journal} {\bibinfo  {journal} {Physics Reports}\ }\textbf {\bibinfo {volume} {480}},\ \bibinfo {pages} {57–159} (\bibinfo {year} {2009})}\BibitemShut {NoStop}%
\bibitem [{\citenamefont {Campa}\ \emph {et~al.}(2014)\citenamefont {Campa}, \citenamefont {Dauxois}, \citenamefont {Fanelli},\ and\ \citenamefont {Ruffo}}]{campa2014physics}%
  \BibitemOpen
  \bibfield  {author} {\bibinfo {author} {\bibfnamefont {A.}~\bibnamefont {Campa}}, \bibinfo {author} {\bibfnamefont {T.}~\bibnamefont {Dauxois}}, \bibinfo {author} {\bibfnamefont {D.}~\bibnamefont {Fanelli}}, \ and\ \bibinfo {author} {\bibfnamefont {S.}~\bibnamefont {Ruffo}},\ }\href@noop {} {\emph {\bibinfo {title} {Physics of long-range interacting systems}}}\ (\bibinfo  {publisher} {OUP Oxford},\ \bibinfo {year} {2014})\BibitemShut {NoStop}%
\bibitem [{\citenamefont {He}(2012)}]{he2012equilibrium}%
  \BibitemOpen
  \bibfield  {author} {\bibinfo {author} {\bibfnamefont {P.}~\bibnamefont {He}},\ }\href {\doibase 10.1111/j.1365‐2966.2011.19830.x} {\bibfield  {journal} {\bibinfo  {journal} {Monthly Notices of the Royal Astronomical Society}\ }\textbf {\bibinfo {volume} {419}},\ \bibinfo {pages} {1667‐1681} (\bibinfo {year} {2012})}\BibitemShut {NoStop}%
\bibitem [{\citenamefont {Defenu}\ \emph {et~al.}(2023)\citenamefont {Defenu}, \citenamefont {Donner}, \citenamefont {Macr\`i}, \citenamefont {Pagano}, \citenamefont {Ruffo},\ and\ \citenamefont {Trombettoni}}]{defenu2023longrange}%
  \BibitemOpen
  \bibfield  {author} {\bibinfo {author} {\bibfnamefont {N.}~\bibnamefont {Defenu}}, \bibinfo {author} {\bibfnamefont {T.}~\bibnamefont {Donner}}, \bibinfo {author} {\bibfnamefont {T.}~\bibnamefont {Macr\`i}}, \bibinfo {author} {\bibfnamefont {G.}~\bibnamefont {Pagano}}, \bibinfo {author} {\bibfnamefont {S.}~\bibnamefont {Ruffo}}, \ and\ \bibinfo {author} {\bibfnamefont {A.}~\bibnamefont {Trombettoni}},\ }\href {\doibase 10.1103/RevModPhys.95.035002} {\bibfield  {journal} {\bibinfo  {journal} {Rev. Mod. Phys.}\ }\textbf {\bibinfo {volume} {95}},\ \bibinfo {pages} {035002} (\bibinfo {year} {2023})}\BibitemShut {NoStop}%
\bibitem [{\citenamefont {Parisi}(1988)}]{parisi1988statistical}%
  \BibitemOpen
  \bibfield  {author} {\bibinfo {author} {\bibfnamefont {G.}~\bibnamefont {Parisi}},\ }\href {https://cds.cern.ch/record/111935} {\emph {\bibinfo {title} {{Statistical field theory}}}},\ Frontiers in physics\ (\bibinfo  {publisher} {Addison-Wesley},\ \bibinfo {address} {Redwood City, CA},\ \bibinfo {year} {1988})\BibitemShut {NoStop}%
\bibitem [{\citenamefont {Nishimori}\ and\ \citenamefont {Ortiz}(2011)}]{nishimori2011elements}%
  \BibitemOpen
  \bibfield  {author} {\bibinfo {author} {\bibfnamefont {H.}~\bibnamefont {Nishimori}}\ and\ \bibinfo {author} {\bibfnamefont {G.}~\bibnamefont {Ortiz}},\ }\href {http://www.loc.gov/catdir/enhancements/fy1108/2010032104-t.html} {\emph {\bibinfo {title} {Elements of phase transitions and critical phenomena}}},\ Oxford graduate texts\ (\bibinfo  {publisher} {Oxford University Press},\ \bibinfo {address} {Oxford},\ \bibinfo {year} {2011})\BibitemShut {NoStop}%
\bibitem [{\citenamefont {Kac}\ \emph {et~al.}(1963)\citenamefont {Kac}, \citenamefont {Uhlenbeck},\ and\ \citenamefont {Hemmer}}]{kac1963van}%
  \BibitemOpen
  \bibfield  {author} {\bibinfo {author} {\bibfnamefont {M.}~\bibnamefont {Kac}}, \bibinfo {author} {\bibfnamefont {G.}~\bibnamefont {Uhlenbeck}}, \ and\ \bibinfo {author} {\bibfnamefont {P.}~\bibnamefont {Hemmer}},\ }\href@noop {} {\bibfield  {journal} {\bibinfo  {journal} {Journal of Mathematical Physics}\ }\textbf {\bibinfo {volume} {4}},\ \bibinfo {pages} {216} (\bibinfo {year} {1963})}\BibitemShut {NoStop}%
\bibitem [{\citenamefont {Mori}(2010)}]{mori2010analysis}%
  \BibitemOpen
  \bibfield  {author} {\bibinfo {author} {\bibfnamefont {T.}~\bibnamefont {Mori}},\ }\href {\doibase 10.1103/PhysRevE.82.060103} {\bibfield  {journal} {\bibinfo  {journal} {Phys. Rev. E}\ }\textbf {\bibinfo {volume} {82}},\ \bibinfo {pages} {060103} (\bibinfo {year} {2010})}\BibitemShut {NoStop}%
\bibitem [{\citenamefont {Mori}(2012)}]{mori2012equilibrium}%
  \BibitemOpen
  \bibfield  {author} {\bibinfo {author} {\bibfnamefont {T.}~\bibnamefont {Mori}},\ }\href {\doibase 10.1103/PhysRevE.86.021132} {\bibfield  {journal} {\bibinfo  {journal} {Phys. Rev. E}\ }\textbf {\bibinfo {volume} {86}},\ \bibinfo {pages} {021132} (\bibinfo {year} {2012})}\BibitemShut {NoStop}%
\bibitem [{\citenamefont {Botzung}\ \emph {et~al.}(2021)\citenamefont {Botzung}, \citenamefont {Hagenm\"uller}, \citenamefont {Masella}, \citenamefont {Dubail}, \citenamefont {Defenu}, \citenamefont {Trombettoni},\ and\ \citenamefont {Pupillo}}]{botzung2021effects}%
  \BibitemOpen
  \bibfield  {author} {\bibinfo {author} {\bibfnamefont {T.}~\bibnamefont {Botzung}}, \bibinfo {author} {\bibfnamefont {D.}~\bibnamefont {Hagenm\"uller}}, \bibinfo {author} {\bibfnamefont {G.}~\bibnamefont {Masella}}, \bibinfo {author} {\bibfnamefont {J.}~\bibnamefont {Dubail}}, \bibinfo {author} {\bibfnamefont {N.}~\bibnamefont {Defenu}}, \bibinfo {author} {\bibfnamefont {A.}~\bibnamefont {Trombettoni}}, \ and\ \bibinfo {author} {\bibfnamefont {G.}~\bibnamefont {Pupillo}},\ }\href {\doibase 10.1103/PhysRevB.103.155139} {\bibfield  {journal} {\bibinfo  {journal} {Phys. Rev. B}\ }\textbf {\bibinfo {volume} {103}},\ \bibinfo {pages} {155139} (\bibinfo {year} {2021})}\BibitemShut {NoStop}%
\bibitem [{\citenamefont {Étienne Granet}(2023)}]{granet2023exact}%
  \BibitemOpen
  \bibfield  {author} {\bibinfo {author} {\bibnamefont {Étienne Granet}},\ }\href {\doibase 10.21468/SciPostPhys.14.5.133} {\bibfield  {journal} {\bibinfo  {journal} {SciPost Phys.}\ }\textbf {\bibinfo {volume} {14}},\ \bibinfo {pages} {133} (\bibinfo {year} {2023})}\BibitemShut {NoStop}%
\bibitem [{\citenamefont {Kastner}(2025)}]{kastner2025long}%
  \BibitemOpen
  \bibfield  {author} {\bibinfo {author} {\bibfnamefont {M.}~\bibnamefont {Kastner}},\ }\href@noop {} {\enquote {\bibinfo {title} {Long-range systems, (non)extensivity, and the rescaling of energies},}\ } (\bibinfo {year} {2025}),\ \Eprint {http://arxiv.org/abs/arxiv:2506.22296} {arxiv:2506.22296} \BibitemShut {NoStop}%
\bibitem [{\citenamefont {Monroe}\ \emph {et~al.}(2021)\citenamefont {Monroe}, \citenamefont {Campbell}, \citenamefont {Duan}, \citenamefont {Gong}, \citenamefont {Gorshkov}, \citenamefont {Hess}, \citenamefont {Islam}, \citenamefont {Kim}, \citenamefont {Linke}, \citenamefont {Pagano}, \citenamefont {Richerme}, \citenamefont {Senko},\ and\ \citenamefont {Yao}}]{monroe2021programmable}%
  \BibitemOpen
  \bibfield  {author} {\bibinfo {author} {\bibfnamefont {C.}~\bibnamefont {Monroe}}, \bibinfo {author} {\bibfnamefont {W.~C.}\ \bibnamefont {Campbell}}, \bibinfo {author} {\bibfnamefont {L.-M.}\ \bibnamefont {Duan}}, \bibinfo {author} {\bibfnamefont {Z.-X.}\ \bibnamefont {Gong}}, \bibinfo {author} {\bibfnamefont {A.~V.}\ \bibnamefont {Gorshkov}}, \bibinfo {author} {\bibfnamefont {P.~W.}\ \bibnamefont {Hess}}, \bibinfo {author} {\bibfnamefont {R.}~\bibnamefont {Islam}}, \bibinfo {author} {\bibfnamefont {K.}~\bibnamefont {Kim}}, \bibinfo {author} {\bibfnamefont {N.~M.}\ \bibnamefont {Linke}}, \bibinfo {author} {\bibfnamefont {G.}~\bibnamefont {Pagano}}, \bibinfo {author} {\bibfnamefont {P.}~\bibnamefont {Richerme}}, \bibinfo {author} {\bibfnamefont {C.}~\bibnamefont {Senko}}, \ and\ \bibinfo {author} {\bibfnamefont {N.~Y.}\ \bibnamefont {Yao}},\ }\href {\doibase 10.1103/RevModPhys.93.025001} {\bibfield  {journal} {\bibinfo  {journal} {Rev. Mod. Phys.}\ }\textbf {\bibinfo {volume} {93}},\ \bibinfo {pages}
  {025001} (\bibinfo {year} {2021})}\BibitemShut {NoStop}%
\bibitem [{\citenamefont {Degen}\ \emph {et~al.}(2017)\citenamefont {Degen}, \citenamefont {Reinhard},\ and\ \citenamefont {Cappellaro}}]{degen2017quantum}%
  \BibitemOpen
  \bibfield  {author} {\bibinfo {author} {\bibfnamefont {C.~L.}\ \bibnamefont {Degen}}, \bibinfo {author} {\bibfnamefont {F.}~\bibnamefont {Reinhard}}, \ and\ \bibinfo {author} {\bibfnamefont {P.}~\bibnamefont {Cappellaro}},\ }\href {\doibase 10.1103/RevModPhys.89.035002} {\bibfield  {journal} {\bibinfo  {journal} {Rev. Mod. Phys.}\ }\textbf {\bibinfo {volume} {89}},\ \bibinfo {pages} {035002} (\bibinfo {year} {2017})}\BibitemShut {NoStop}%
\bibitem [{\citenamefont {Pezze}\ \emph {et~al.}(2018)\citenamefont {Pezze}, \citenamefont {Smerzi}, \citenamefont {Oberthaler}, \citenamefont {Schmied},\ and\ \citenamefont {Treutlein}}]{pezze2018quantum}%
  \BibitemOpen
  \bibfield  {author} {\bibinfo {author} {\bibfnamefont {L.}~\bibnamefont {Pezze}}, \bibinfo {author} {\bibfnamefont {A.}~\bibnamefont {Smerzi}}, \bibinfo {author} {\bibfnamefont {M.~K.}\ \bibnamefont {Oberthaler}}, \bibinfo {author} {\bibfnamefont {R.}~\bibnamefont {Schmied}}, \ and\ \bibinfo {author} {\bibfnamefont {P.}~\bibnamefont {Treutlein}},\ }\href@noop {} {\bibfield  {journal} {\bibinfo  {journal} {Reviews of Modern Physics}\ }\textbf {\bibinfo {volume} {90}},\ \bibinfo {pages} {035005} (\bibinfo {year} {2018})}\BibitemShut {NoStop}%
\bibitem [{\citenamefont {Giovannetti}\ \emph {et~al.}(2004)\citenamefont {Giovannetti}, \citenamefont {Lloyd},\ and\ \citenamefont {Maccone}}]{giovannetti2004quantum}%
  \BibitemOpen
  \bibfield  {author} {\bibinfo {author} {\bibfnamefont {V.}~\bibnamefont {Giovannetti}}, \bibinfo {author} {\bibfnamefont {S.}~\bibnamefont {Lloyd}}, \ and\ \bibinfo {author} {\bibfnamefont {L.}~\bibnamefont {Maccone}},\ }\href {\doibase 10.1126/science.1104149} {\bibfield  {journal} {\bibinfo  {journal} {Science}\ }\textbf {\bibinfo {volume} {306}},\ \bibinfo {pages} {1330} (\bibinfo {year} {2004})}\BibitemShut {NoStop}%
\bibitem [{\citenamefont {Demkowicz-Dobrza{\'{n}}ski}\ \emph {et~al.}(2012)\citenamefont {Demkowicz-Dobrza{\'{n}}ski}, \citenamefont {Ko{\l}ody{\'{n}}ski},\ and\ \citenamefont {Gu{\c{T}}{\u{a}}}}]{demkowiczdobrzański2012elusive}%
  \BibitemOpen
  \bibfield  {author} {\bibinfo {author} {\bibfnamefont {R.}~\bibnamefont {Demkowicz-Dobrza{\'{n}}ski}}, \bibinfo {author} {\bibfnamefont {J.}~\bibnamefont {Ko{\l}ody{\'{n}}ski}}, \ and\ \bibinfo {author} {\bibfnamefont {M.}~\bibnamefont {Gu{\c{T}}{\u{a}}}},\ }\href {\doibase 10.1038/ncomms2067} {\bibfield  {journal} {\bibinfo  {journal} {Nature Communications}\ }\textbf {\bibinfo {volume} {3}},\ \bibinfo {pages} {1063} (\bibinfo {year} {2012})}\BibitemShut {NoStop}%
\bibitem [{\citenamefont {Wolswijk}\ \emph {et~al.}(2025)\citenamefont {Wolswijk}, \citenamefont {Cavicchioli}, \citenamefont {Vinelli}, \citenamefont {Chiarotti}, \citenamefont {Donati}, \citenamefont {Fernandez}, \citenamefont {Rajkov}, \citenamefont {Mancini}, \citenamefont {Vezio}, \citenamefont {Zhou}, \citenamefont {Pace}, \citenamefont {Mazzinghi}, \citenamefont {Antolini}, \citenamefont {Salvi},\ and\ \citenamefont {Gavryusev}}]{wolswijk2025trapping}%
  \BibitemOpen
  \bibfield  {author} {\bibinfo {author} {\bibfnamefont {L.}~\bibnamefont {Wolswijk}}, \bibinfo {author} {\bibfnamefont {L.}~\bibnamefont {Cavicchioli}}, \bibinfo {author} {\bibfnamefont {G.}~\bibnamefont {Vinelli}}, \bibinfo {author} {\bibfnamefont {M.}~\bibnamefont {Chiarotti}}, \bibinfo {author} {\bibfnamefont {L.}~\bibnamefont {Donati}}, \bibinfo {author} {\bibfnamefont {M.~F.}\ \bibnamefont {Fernandez}}, \bibinfo {author} {\bibfnamefont {D.~H.}\ \bibnamefont {Rajkov}}, \bibinfo {author} {\bibfnamefont {C.}~\bibnamefont {Mancini}}, \bibinfo {author} {\bibfnamefont {P.}~\bibnamefont {Vezio}}, \bibinfo {author} {\bibfnamefont {T.}~\bibnamefont {Zhou}}, \bibinfo {author} {\bibfnamefont {G.~D.}\ \bibnamefont {Pace}}, \bibinfo {author} {\bibfnamefont {C.}~\bibnamefont {Mazzinghi}}, \bibinfo {author} {\bibfnamefont {N.}~\bibnamefont {Antolini}}, \bibinfo {author} {\bibfnamefont {L.}~\bibnamefont {Salvi}}, \ and\ \bibinfo {author} {\bibfnamefont {V.}~\bibnamefont {Gavryusev}},\ }\href@noop {} {\enquote {\bibinfo
  {title} {Trapping, manipulating and probing ultracold atoms: a quantum technologies tutorial},}\ } (\bibinfo {year} {2025}),\ \Eprint {http://arxiv.org/abs/arXiv:2510.20790} {arXiv:2510.20790} \BibitemShut {NoStop}%
\bibitem [{\citenamefont {Demkowicz-Dobrzanski}\ \emph {et~al.}(2009)\citenamefont {Demkowicz-Dobrzanski}, \citenamefont {Dorner}, \citenamefont {Smith}, \citenamefont {Lundeen}, \citenamefont {Wasilewski}, \citenamefont {Banaszek},\ and\ \citenamefont {Walmsley}}]{demkowiczdobrzanski2009quantum}%
  \BibitemOpen
  \bibfield  {author} {\bibinfo {author} {\bibfnamefont {R.}~\bibnamefont {Demkowicz-Dobrzanski}}, \bibinfo {author} {\bibfnamefont {U.}~\bibnamefont {Dorner}}, \bibinfo {author} {\bibfnamefont {B.~J.}\ \bibnamefont {Smith}}, \bibinfo {author} {\bibfnamefont {J.~S.}\ \bibnamefont {Lundeen}}, \bibinfo {author} {\bibfnamefont {W.}~\bibnamefont {Wasilewski}}, \bibinfo {author} {\bibfnamefont {K.}~\bibnamefont {Banaszek}}, \ and\ \bibinfo {author} {\bibfnamefont {I.~A.}\ \bibnamefont {Walmsley}},\ }\href {\doibase 10.1103/PhysRevA.80.013825} {\bibfield  {journal} {\bibinfo  {journal} {Phys. Rev. A}\ }\textbf {\bibinfo {volume} {80}},\ \bibinfo {pages} {013825} (\bibinfo {year} {2009})}\BibitemShut {NoStop}%
\bibitem [{\citenamefont {Escher}\ \emph {et~al.}(2011)\citenamefont {Escher}, \citenamefont {de~Matos~Filho},\ and\ \citenamefont {Davidovich}}]{escher2011general}%
  \BibitemOpen
  \bibfield  {author} {\bibinfo {author} {\bibfnamefont {B.}~\bibnamefont {Escher}}, \bibinfo {author} {\bibfnamefont {R.~L.}\ \bibnamefont {de~Matos~Filho}}, \ and\ \bibinfo {author} {\bibfnamefont {L.}~\bibnamefont {Davidovich}},\ }\href@noop {} {\bibfield  {journal} {\bibinfo  {journal} {Nature Physics}\ }\textbf {\bibinfo {volume} {7}},\ \bibinfo {pages} {406} (\bibinfo {year} {2011})}\BibitemShut {NoStop}%
\bibitem [{\citenamefont {Escher}\ \emph {et~al.}(2012)\citenamefont {Escher}, \citenamefont {Davidovich}, \citenamefont {Zagury},\ and\ \citenamefont {de~Matos~Filho}}]{escher2012quantum}%
  \BibitemOpen
  \bibfield  {author} {\bibinfo {author} {\bibfnamefont {B.~M.}\ \bibnamefont {Escher}}, \bibinfo {author} {\bibfnamefont {L.}~\bibnamefont {Davidovich}}, \bibinfo {author} {\bibfnamefont {N.}~\bibnamefont {Zagury}}, \ and\ \bibinfo {author} {\bibfnamefont {R.~L.}\ \bibnamefont {de~Matos~Filho}},\ }\href {\doibase 10.1103/PhysRevLett.109.190404} {\bibfield  {journal} {\bibinfo  {journal} {Phys. Rev. Lett.}\ }\textbf {\bibinfo {volume} {109}},\ \bibinfo {pages} {190404} (\bibinfo {year} {2012})}\BibitemShut {NoStop}%
\bibitem [{\citenamefont {Das}\ \emph {et~al.}(2025)\citenamefont {Das}, \citenamefont {G\'orecki},\ and\ \citenamefont {Demkowicz-Dobrza\ifmmode~\acute{n}\else \'{n}\fi{}ski}}]{das2012universal}%
  \BibitemOpen
  \bibfield  {author} {\bibinfo {author} {\bibfnamefont {A.}~\bibnamefont {Das}}, \bibinfo {author} {\bibfnamefont {W.}~\bibnamefont {G\'orecki}}, \ and\ \bibinfo {author} {\bibfnamefont {R.}~\bibnamefont {Demkowicz-Dobrza\ifmmode~\acute{n}\else \'{n}\fi{}ski}},\ }\href {\doibase 10.1103/PhysRevA.111.L020403} {\bibfield  {journal} {\bibinfo  {journal} {Phys. Rev. A}\ }\textbf {\bibinfo {volume} {111}},\ \bibinfo {pages} {L020403} (\bibinfo {year} {2025})}\BibitemShut {NoStop}%
\bibitem [{\citenamefont {Demkowicz-Dobrza\ifmmode~\acute{n}\else \'{n}\fi{}ski}\ and\ \citenamefont {Maccone}(2014)}]{demkowiczdobrzanski2014using}%
  \BibitemOpen
  \bibfield  {author} {\bibinfo {author} {\bibfnamefont {R.}~\bibnamefont {Demkowicz-Dobrza\ifmmode~\acute{n}\else \'{n}\fi{}ski}}\ and\ \bibinfo {author} {\bibfnamefont {L.}~\bibnamefont {Maccone}},\ }\href {\doibase 10.1103/PhysRevLett.113.250801} {\bibfield  {journal} {\bibinfo  {journal} {Phys. Rev. Lett.}\ }\textbf {\bibinfo {volume} {113}},\ \bibinfo {pages} {250801} (\bibinfo {year} {2014})}\BibitemShut {NoStop}%
\bibitem [{\citenamefont {Yue}\ \emph {et~al.}(2014)\citenamefont {Yue}, \citenamefont {Zhang},\ and\ \citenamefont {Fan}}]{yue2014quantum}%
  \BibitemOpen
  \bibfield  {author} {\bibinfo {author} {\bibfnamefont {J.-D.}\ \bibnamefont {Yue}}, \bibinfo {author} {\bibfnamefont {Y.-R.}\ \bibnamefont {Zhang}}, \ and\ \bibinfo {author} {\bibfnamefont {H.}~\bibnamefont {Fan}},\ }\href {\doibase 10.1038/srep05933} {\bibfield  {journal} {\bibinfo  {journal} {Scientific Reports}\ }\textbf {\bibinfo {volume} {4}},\ \bibinfo {pages} {5933} (\bibinfo {year} {2014})}\BibitemShut {NoStop}%
\bibitem [{\citenamefont {Ko{\l}ody{\'n}ski}(2014)}]{kolodynski2014precision}%
  \BibitemOpen
  \bibfield  {author} {\bibinfo {author} {\bibfnamefont {J.}~\bibnamefont {Ko{\l}ody{\'n}ski}},\ }\emph {\bibinfo {title} {Precision bounds in noisy quantum metrology}},\ \href@noop {} {Ph.D. thesis},\ \bibinfo  {school} {University of Warsaw} (\bibinfo {year} {2014}),\ \Eprint {http://arxiv.org/abs/arXiv:1409.0535v2} {arXiv:1409.0535v2} \BibitemShut {NoStop}%
\bibitem [{\citenamefont {Demkowicz-Dobrza\ifmmode~\acute{n}\else \'{n}\fi{}ski}\ \emph {et~al.}(2017)\citenamefont {Demkowicz-Dobrza\ifmmode~\acute{n}\else \'{n}\fi{}ski}, \citenamefont {Czajkowski},\ and\ \citenamefont {Sekatski}}]{demkowiczdobrzanski2017adaptive}%
  \BibitemOpen
  \bibfield  {author} {\bibinfo {author} {\bibfnamefont {R.}~\bibnamefont {Demkowicz-Dobrza\ifmmode~\acute{n}\else \'{n}\fi{}ski}}, \bibinfo {author} {\bibfnamefont {J.}~\bibnamefont {Czajkowski}}, \ and\ \bibinfo {author} {\bibfnamefont {P.}~\bibnamefont {Sekatski}},\ }\href {\doibase 10.1103/PhysRevX.7.041009} {\bibfield  {journal} {\bibinfo  {journal} {Phys. Rev. X}\ }\textbf {\bibinfo {volume} {7}},\ \bibinfo {pages} {041009} (\bibinfo {year} {2017})}\BibitemShut {NoStop}%
\bibitem [{\citenamefont {Albarelli}\ \emph {et~al.}(2018)\citenamefont {Albarelli}, \citenamefont {Rossi}, \citenamefont {Tamascelli},\ and\ \citenamefont {Genoni}}]{albarelli2018restoring}%
  \BibitemOpen
  \bibfield  {author} {\bibinfo {author} {\bibfnamefont {F.}~\bibnamefont {Albarelli}}, \bibinfo {author} {\bibfnamefont {M.~A.~C.}\ \bibnamefont {Rossi}}, \bibinfo {author} {\bibfnamefont {D.}~\bibnamefont {Tamascelli}}, \ and\ \bibinfo {author} {\bibfnamefont {M.~G.}\ \bibnamefont {Genoni}},\ }\href {\doibase 10.22331/q-2018-12-03-110} {\bibfield  {journal} {\bibinfo  {journal} {Quantum}\ }\textbf {\bibinfo {volume} {2}},\ \bibinfo {pages} {110} (\bibinfo {year} {2018})}\BibitemShut {NoStop}%
\bibitem [{\citenamefont {Albarelli}\ \emph {et~al.}(2022)\citenamefont {Albarelli}, \citenamefont {Mazelanik}, \citenamefont {Lipka}, \citenamefont {Streltsov}, \citenamefont {Parniak},\ and\ \citenamefont {Demkowicz-Dobrza\ifmmode~\acute{n}\else \'{n}\fi{}ski}}]{albarelli2022quantum}%
  \BibitemOpen
  \bibfield  {author} {\bibinfo {author} {\bibfnamefont {F.}~\bibnamefont {Albarelli}}, \bibinfo {author} {\bibfnamefont {M.}~\bibnamefont {Mazelanik}}, \bibinfo {author} {\bibfnamefont {M.}~\bibnamefont {Lipka}}, \bibinfo {author} {\bibfnamefont {A.}~\bibnamefont {Streltsov}}, \bibinfo {author} {\bibfnamefont {M.}~\bibnamefont {Parniak}}, \ and\ \bibinfo {author} {\bibfnamefont {R.}~\bibnamefont {Demkowicz-Dobrza\ifmmode~\acute{n}\else \'{n}\fi{}ski}},\ }\href {\doibase 10.1103/PhysRevLett.128.240504} {\bibfield  {journal} {\bibinfo  {journal} {Phys. Rev. Lett.}\ }\textbf {\bibinfo {volume} {128}},\ \bibinfo {pages} {240504} (\bibinfo {year} {2022})}\BibitemShut {NoStop}%
\bibitem [{\citenamefont {G\'orecki}\ \emph {et~al.}(2022)\citenamefont {G\'orecki}, \citenamefont {Riccardi},\ and\ \citenamefont {Maccone}}]{gorecki2022quantum}%
  \BibitemOpen
  \bibfield  {author} {\bibinfo {author} {\bibfnamefont {W.}~\bibnamefont {G\'orecki}}, \bibinfo {author} {\bibfnamefont {A.}~\bibnamefont {Riccardi}}, \ and\ \bibinfo {author} {\bibfnamefont {L.}~\bibnamefont {Maccone}},\ }\href {\doibase 10.1103/PhysRevLett.129.240503} {\bibfield  {journal} {\bibinfo  {journal} {Phys. Rev. Lett.}\ }\textbf {\bibinfo {volume} {129}},\ \bibinfo {pages} {240503} (\bibinfo {year} {2022})}\BibitemShut {NoStop}%
\bibitem [{\citenamefont {Bao}\ \emph {et~al.}(2022)\citenamefont {Bao}, \citenamefont {Qi}, \citenamefont {Wang}, \citenamefont {Dong},\ and\ \citenamefont {Wu}}]{bao2022multichannel}%
  \BibitemOpen
  \bibfield  {author} {\bibinfo {author} {\bibfnamefont {L.}~\bibnamefont {Bao}}, \bibinfo {author} {\bibfnamefont {B.}~\bibnamefont {Qi}}, \bibinfo {author} {\bibfnamefont {Y.}~\bibnamefont {Wang}}, \bibinfo {author} {\bibfnamefont {D.}~\bibnamefont {Dong}}, \ and\ \bibinfo {author} {\bibfnamefont {R.}~\bibnamefont {Wu}},\ }\href {\doibase 10.1007/s11432-020-3196-x} {\bibfield  {journal} {\bibinfo  {journal} {Science China Information Sciences}\ }\textbf {\bibinfo {volume} {65}},\ \bibinfo {pages} {200505} (\bibinfo {year} {2022})}\BibitemShut {NoStop}%
\bibitem [{\citenamefont {G{\'o}recki}\ \emph {et~al.}(2025)\citenamefont {G{\'o}recki}, \citenamefont {Albarelli}, \citenamefont {Felicetti}, \citenamefont {Di~Candia},\ and\ \citenamefont {Maccone}}]{gorecki2025interplay}%
  \BibitemOpen
  \bibfield  {author} {\bibinfo {author} {\bibfnamefont {W.}~\bibnamefont {G{\'o}recki}}, \bibinfo {author} {\bibfnamefont {F.}~\bibnamefont {Albarelli}}, \bibinfo {author} {\bibfnamefont {S.}~\bibnamefont {Felicetti}}, \bibinfo {author} {\bibfnamefont {R.}~\bibnamefont {Di~Candia}}, \ and\ \bibinfo {author} {\bibfnamefont {L.}~\bibnamefont {Maccone}},\ }\href@noop {} {\bibfield  {journal} {\bibinfo  {journal} {PRX Quantum}\ }\textbf {\bibinfo {volume} {6}},\ \bibinfo {pages} {020351} (\bibinfo {year} {2025})}\BibitemShut {NoStop}%
\bibitem [{\citenamefont {Das}\ and\ \citenamefont {Demkowicz-Dobrzański}(2025)}]{das2025quantum}%
  \BibitemOpen
  \bibfield  {author} {\bibinfo {author} {\bibfnamefont {A.}~\bibnamefont {Das}}\ and\ \bibinfo {author} {\bibfnamefont {R.}~\bibnamefont {Demkowicz-Dobrzański}},\ }\href@noop {} {\enquote {\bibinfo {title} {Quantum metrology in presence of correlated noise via markovian embedding},}\ } (\bibinfo {year} {2025}),\ \Eprint {http://arxiv.org/abs/arxiv:2509.19685} {arxiv:2509.19685} \BibitemShut {NoStop}%
\bibitem [{\citenamefont {Kurdzia\l{}ek}\ \emph {et~al.}(2025)\citenamefont {Kurdzia\l{}ek}, \citenamefont {Albarelli},\ and\ \citenamefont {Demkowicz-Dobrza\ifmmode~\acute{n}\else \'{n}\fi{}ski}}]{kurdzialek2025universal}%
  \BibitemOpen
  \bibfield  {author} {\bibinfo {author} {\bibfnamefont {S.}~\bibnamefont {Kurdzia\l{}ek}}, \bibinfo {author} {\bibfnamefont {F.}~\bibnamefont {Albarelli}}, \ and\ \bibinfo {author} {\bibfnamefont {R.}~\bibnamefont {Demkowicz-Dobrza\ifmmode~\acute{n}\else \'{n}\fi{}ski}},\ }\href {\doibase 10.1103/jy3v-wkcb} {\bibfield  {journal} {\bibinfo  {journal} {Phys. Rev. Lett.}\ }\textbf {\bibinfo {volume} {135}},\ \bibinfo {pages} {130801} (\bibinfo {year} {2025})}\BibitemShut {NoStop}%
\bibitem [{\citenamefont {Breuer}\ and\ \citenamefont {Petruccione}(2002)}]{breuer2002theory}%
  \BibitemOpen
  \bibfield  {author} {\bibinfo {author} {\bibfnamefont {H.-P.}\ \bibnamefont {Breuer}}\ and\ \bibinfo {author} {\bibfnamefont {F.}~\bibnamefont {Petruccione}},\ }\href@noop {} {\emph {\bibinfo {title} {The theory of open quantum systems}}}\ (\bibinfo  {publisher} {OUP Oxford},\ \bibinfo {year} {2002})\BibitemShut {NoStop}%
\bibitem [{\citenamefont {Caldeira}\ and\ \citenamefont {J. Leggett}(1981)}]{caldeira1981influence}%
  \BibitemOpen
  \bibfield  {author} {\bibinfo {author} {\bibfnamefont {A.~O.}\ \bibnamefont {Caldeira}}\ and\ \bibinfo {author} {\bibfnamefont {A.}~\bibnamefont {J. Leggett}},\ }\href {\doibase 10.1103/PhysRevLett.46.211} {\bibfield  {journal} {\bibinfo  {journal} {Physical Review Letters}\ }\textbf {\bibinfo {volume} {46}},\ \bibinfo {pages} {211} (\bibinfo {year} {1981})}\BibitemShut {NoStop}%
\bibitem [{\citenamefont {Leggett}\ \emph {et~al.}(1987)\citenamefont {Leggett}, \citenamefont {Chakravarty}, \citenamefont {Dorsey}, \citenamefont {Fisher}, \citenamefont {Garg},\ and\ \citenamefont {Zwerger}}]{leggett1987dynamics}%
  \BibitemOpen
  \bibfield  {author} {\bibinfo {author} {\bibfnamefont {A.~J.}\ \bibnamefont {Leggett}}, \bibinfo {author} {\bibfnamefont {S.}~\bibnamefont {Chakravarty}}, \bibinfo {author} {\bibfnamefont {A.~T.}\ \bibnamefont {Dorsey}}, \bibinfo {author} {\bibfnamefont {M.~P.~A.}\ \bibnamefont {Fisher}}, \bibinfo {author} {\bibfnamefont {A.}~\bibnamefont {Garg}}, \ and\ \bibinfo {author} {\bibfnamefont {W.}~\bibnamefont {Zwerger}},\ }\href {\doibase 10.1103/RevModPhys.59.1} {\bibfield  {journal} {\bibinfo  {journal} {Rev. Mod. Phys.}\ }\textbf {\bibinfo {volume} {59}},\ \bibinfo {pages} {1} (\bibinfo {year} {1987})}\BibitemShut {NoStop}%
\bibitem [{\citenamefont {Javanainen}\ and\ \citenamefont {Yoo}(1996)}]{javanainen1996quantum}%
  \BibitemOpen
  \bibfield  {author} {\bibinfo {author} {\bibfnamefont {J.}~\bibnamefont {Javanainen}}\ and\ \bibinfo {author} {\bibfnamefont {S.~M.}\ \bibnamefont {Yoo}},\ }\href {\doibase 10.1103/PhysRevLett.76.161} {\bibfield  {journal} {\bibinfo  {journal} {Phys. Rev. Lett.}\ }\textbf {\bibinfo {volume} {76}},\ \bibinfo {pages} {161} (\bibinfo {year} {1996})}\BibitemShut {NoStop}%
\bibitem [{\citenamefont {Caves}(1981)}]{caves1981quantum}%
  \BibitemOpen
  \bibfield  {author} {\bibinfo {author} {\bibfnamefont {C.~M.}\ \bibnamefont {Caves}},\ }\href@noop {} {\bibfield  {journal} {\bibinfo  {journal} {Physical Review D}\ }\textbf {\bibinfo {volume} {23}},\ \bibinfo {pages} {1693} (\bibinfo {year} {1981})}\BibitemShut {NoStop}%
\bibitem [{\citenamefont {Pezz{\'e}}\ and\ \citenamefont {Smerzi}(2009)}]{pezze2009entanglement}%
  \BibitemOpen
  \bibfield  {author} {\bibinfo {author} {\bibfnamefont {L.}~\bibnamefont {Pezz{\'e}}}\ and\ \bibinfo {author} {\bibfnamefont {A.}~\bibnamefont {Smerzi}},\ }\href@noop {} {\bibfield  {journal} {\bibinfo  {journal} {Physical review letters}\ }\textbf {\bibinfo {volume} {102}},\ \bibinfo {pages} {100401} (\bibinfo {year} {2009})}\BibitemShut {NoStop}%
\bibitem [{\citenamefont {Braunstein}\ and\ \citenamefont {Caves}(1994)}]{braunstein1994statistical}%
  \BibitemOpen
  \bibfield  {author} {\bibinfo {author} {\bibfnamefont {S.~L.}\ \bibnamefont {Braunstein}}\ and\ \bibinfo {author} {\bibfnamefont {C.~M.}\ \bibnamefont {Caves}},\ }\href@noop {} {\bibfield  {journal} {\bibinfo  {journal} {Physical Review Letters}\ }\textbf {\bibinfo {volume} {72}},\ \bibinfo {pages} {3439} (\bibinfo {year} {1994})}\BibitemShut {NoStop}%
\bibitem [{\citenamefont {Louisell}\ and\ \citenamefont {Marburger}(1967)}]{louisell1967solutions}%
  \BibitemOpen
  \bibfield  {author} {\bibinfo {author} {\bibfnamefont {W.}~\bibnamefont {Louisell}}\ and\ \bibinfo {author} {\bibfnamefont {J.}~\bibnamefont {Marburger}},\ }\href {\doibase 10.1109/JQE.1967.1074605} {\bibfield  {journal} {\bibinfo  {journal} {IEEE Journal of Quantum Electronics}\ }\textbf {\bibinfo {volume} {3}},\ \bibinfo {pages} {348} (\bibinfo {year} {1967})}\BibitemShut {NoStop}%
\bibitem [{\citenamefont {Schlosshauer}(2019)}]{schlosshauer2019quantum}%
  \BibitemOpen
  \bibfield  {author} {\bibinfo {author} {\bibfnamefont {M.}~\bibnamefont {Schlosshauer}},\ }\href@noop {} {\bibfield  {journal} {\bibinfo  {journal} {Physics Reports}\ }\textbf {\bibinfo {volume} {831}},\ \bibinfo {pages} {1} (\bibinfo {year} {2019})}\BibitemShut {NoStop}%
\bibitem [{\citenamefont {Qvarfort}\ and\ \citenamefont {Pikovski}(2025)}]{qvarfort2025solving}%
  \BibitemOpen
  \bibfield  {author} {\bibinfo {author} {\bibfnamefont {S.}~\bibnamefont {Qvarfort}}\ and\ \bibinfo {author} {\bibfnamefont {I.}~\bibnamefont {Pikovski}},\ }\href {\doibase 10.1103/PRXQuantum.6.010201} {\bibfield  {journal} {\bibinfo  {journal} {PRX Quantum}\ }\textbf {\bibinfo {volume} {6}},\ \bibinfo {pages} {010201} (\bibinfo {year} {2025})}\BibitemShut {NoStop}%
\bibitem [{\citenamefont {Schaller}(2014)}]{schaller2014open}%
  \BibitemOpen
  \bibfield  {author} {\bibinfo {author} {\bibfnamefont {G.}~\bibnamefont {Schaller}},\ }\href {\doibase 10.1007/978-3-319-03877-3} {\emph {\bibinfo {title} {Open Quantum Systems Far from Equilibrium}}},\ \bibinfo {series} {Lecture Notes in Physics}, Vol.\ \bibinfo {volume} {881}\ (\bibinfo  {publisher} {Springer},\ \bibinfo {address} {Berlin},\ \bibinfo {year} {2014})\BibitemShut {NoStop}%
\end{thebibliography}%

\appendix

\section{Appendix A: Quantum Fisher Information}
    \label{appendix:A5}

In frequentist probability theory, the Fisher Information is a way to quantify the amount of information that a random variable $\vec{a}$ carries about an unknown parameter $\theta$ that characterizes the probability distribution from which $\vec{a}$ is being sampled. It is defined as the variance of the partial derivative with respect to $\theta$ of the natural logarithm of the likelihood function of a model described by $\theta$. Under certain regularity conditions it can be written in a simpler form:

\begin{equation}
    F(\theta) = \sum_{\vec{a}} \frac{1}{P(\vec{a} \mid \theta)} \left( \frac{\partial P(\vec{a} \mid \theta)}{\partial \theta} \right)^{2},
    \label{eq:placeholder_label}
\end{equation}

Where $P(\vec{a} \mid \theta)$ is the probability of sampling $\vec
{a}$, if the probability distribution were to be governed by $\theta$. In parameter estimation theory it provides us with the lower bound on the estimation of an unknown parameter $\theta$ by sampling the random variables $\vec{a}$ from a distribution governed by $\theta$. It is called the Cram\'{e}r-Rao lower bound (CRLB), and it reads:
\begin{equation}
    \Delta \tilde \theta \geq \Delta \theta_{\mathrm{CR}} = \frac{1}{\sqrt{N \cdot F(\theta)}},
    \label{eq:crlb1}
\end{equation}
where $ \Delta\tilde\theta$ denotes the variance of an unbiased estimator $ \tilde\theta$ of $ \theta$ and $N$ the number of times we repeat the measurement.  

In quantum metrology, those concepts have been extended to closed quantum-mechanical systems by \cite{caves1981quantum} to produce the quantum Fisher information and the quantum CRLB. The former is defined as the maximum of the Fisher information over all possible POVM measurements $\mathcal{E}$: 
\begin{equation}
    F_{\mathcal{Q}}\bigl[\hat{\rho}_{\theta}\bigr] = \max_{\mathcal{E}} F(\theta).
    \label{eq:quantum_fisher}
\end{equation}
The quantum CRLB is calculated in the same way by substituting $F_{\mathcal{Q}}\bigl[\hat{\rho}_{\theta}\bigr]$ in (\ref{eq:crlb1}). It can be shown that the quantum CRLB for any separable state of $N$ qubits is lower bounded by 
\begin{equation}
    \Delta \theta_{\text{SQL}} = \frac{1}{ \sqrt{N}},
    \label{eq:SQL}
\end{equation}
which is called the Standard Quantum Limit. On the other hand, for entangled states, it is lower bounded by:
\begin{equation}
    \Delta \theta_{\text{HL}} = \frac{1}{ {N}},
    \label{eq:Heisenberg}
\end{equation}
Also known as the Heisenberg Limit.
Quantum Fisher information and CRLB can also be defined for open quantum systems in terms of Kraus operators describing the dynamics \cite{escher2011general}, or using the final density matrix of the system \cite{pezze2009entanglement} $\hat{\rho}(\theta)={\sum}_{k} \, p_{k}(\theta)|k(\theta)\rangle\langle k(\theta)|$ :
\begin{equation}
    F_{Q}[\hat{\rho}(\theta)]=\sum_{k} \frac{\left(\partial_{\theta} p_{k}\right)^{2}}{p_{k}}+2 \sum_{k, k^{\prime}} \frac{\left(p_{k}-p_{k^{\prime}}\right)^{2}}{p_{k}+p_{k^{\prime}}}\left|\left\langle\partial_{\theta} k \mid k^{\prime}\right\rangle\right|^{2}.
    \label{eq:pezze_fischer}
\end{equation}
In this paper, we use (\ref{eq:pezze_fischer}) on the simulational results for the open quantum interferometer to obtain the quantum Fisher information and the CRLB.

\section{Appendix B: \texorpdfstring{$1$}{1} Harmonic Oscillator and \texorpdfstring{$1$}{1} Harmonic Oscillator in the Environment}
\label{appendix:A1}
Here we present the calculation for the scaling of the interaction energy of the Hamiltonian $H =\omega_a a^{\dagger} a + \omega_R R^{\dagger} R + g(a^{\dagger} R + R^{\dagger} a )$ with the condition  $\langle GS|a^{\dagger} a|GS \rangle =N_A$. This condition can be included in $H$ as $H =\omega_a a^{\dagger} a + \omega_R R^{\dagger} R + g(a^{\dagger} R + R^{\dagger} a )-\mu(N_A -a^{\dagger} a)$ and demanding that $\partial_{\mu} \langle GS| H|GS \rangle = 0$. 
The diagonalization of the Hamiltonian yields:
\begin{equation}
\begin{split}
H &=\omega_a a^{\dagger} a + \omega_R R^{\dagger} R + g(a^{\dagger} R + R^{\dagger} a )-\mu(N_A -a^{\dagger} a)=
\mqty(a^{\dagger} & R^{\dagger})\mqty(\omega_A +\mu& g \\ g & \omega_R)\mqty(a \\ R)-\mu N_A
\\
&=
\mqty(a^{\dagger} & R^{\dagger}) S^{\dagger} \left( \mqty(-\lambda& 0 \\ 0 & +\lambda) + 
\frac{\omega_A+\mu+\omega_R}{2} \mqty(1& 0 \\ 0 & 1) \right) 
S \mqty(a \\ R)-\mu N_A
\\
&=
\mqty(\alpha_0^{\dagger} & \alpha_1^{\dagger})  \left( \mqty(-\lambda& 0 \\ 0 & +\lambda) + \frac{\omega_A+\mu+\omega_R}{2} \mqty(1& 0 \\ 0 & 1) \right)
\mqty(\alpha_0 \\ \alpha_1 )-\mu N_A
\\
&=
\mqty(\alpha_0^{\dagger} & \alpha_1^{\dagger})   \mqty(-\lambda& 0 \\ 0 & +\lambda)  \mqty(\alpha_0 \\ \alpha_1 ) -\mu N_A + \frac{\omega_A+\mu+\omega_R}{2}N
\end{split}
\end{equation}
, where $\lambda = \sqrt{\left( \omega_A + \mu - \omega_R \right)^2 /4+g^2}$, and $S$ is the unitary transformation matrix that diagonalizes the Hamiltonian. We also defined the ladder operators in the new basis as  $ \mqty(\alpha_0 \\  \alpha_1) \equiv S \mqty(a \\ R) $. Notice that the particle number operator
\begin{equation}
\begin{split}
N=a^{\dagger} a + R^{\dagger} R = 
\mqty(a^{\dagger} & R^{\dagger})\mqty(a \\ R)=
\mqty(a^{\dagger} & R^{\dagger})S^{\dagger}S\mqty(a \\ R)=
\mqty(\alpha_0^{\dagger} & \alpha_1^{\dagger})    \mqty(\alpha_0 \\ \alpha_1 )
\end{split}
\end{equation}
commutes with the Hamiltonian $H$. The ground state can thus be identified as $|GS\rangle = \frac{\left( \alpha_0^{\dagger} \right)^N }{\sqrt{N!}} |0\rangle $ with the energy $E_{GS} = -N\lambda(\mu)  -\mu N_A + \frac{\omega_A+\mu+\omega_R}{2}N $, where $\mu$ is determined by the condition $\partial_{\mu} \langle GS| H|GS \rangle = 0$. Calculating the ground state interaction energy using $\mqty(a \\ R) =S^{\dagger} \mqty(\alpha_0 \\  \alpha_1) $ yields:
\begin{equation}
\langle GS| H_I |GS\rangle =\langle GS| g(a^{\dagger} R + R^{\dagger} a )|GS\rangle = 2g \sqrt{N_A N_R} 
\end{equation}
\section{Appendix C: \texorpdfstring{$1$}{1} Harmonic Oscillator and \texorpdfstring{$M$}{1} Harmonic Oscillators in the Environment}
\label{appendix:A2}
We can generalize the procedure of Appendix B to $M$ harmonic oscillators in the environment. The Hamiltonian is thus:
$H =\omega_a a^{\dagger} a + \sum_{i=1}^{M} \omega_i R^{\dagger}_i R_i + \sum_{i=1}^{M} g_i (R^{\dagger}_i a  + a^{\dagger} R_i)$ with the set of conditions  $\langle GS|a^{\dagger} a|GS \rangle =N_A$ and  $\langle GS|R_i^{\dagger} R_i|GS \rangle =M_i$.  We can write $H$ in the matrix notation as:
\begin{equation}
\begin{split}
	H
	&
	=\mqty(a^{\dagger} & R^{\dagger}_1& R^{\dagger}_2&...& R^{\dagger}_N)\mqty(\omega_A& g&g&...&g \\ g & \omega_1&0&...&0\\g&0&\omega_2&...&0\\...&...\\g&0&0&...&\omega_M)\mqty(a \\ R_1\\ R_2\\ ...\\R_N) \equiv
	\\
	&
	\equiv 
	\mqty(\alpha^{\dagger}_0 & \alpha^{\dagger}_1& \alpha^{\dagger}_2&...& \alpha^{\dagger}_N)\mqty(\dmat{\lambda_0,\lambda_1,\lambda_2,..., \lambda_N})	\mqty(\alpha_0\\ \alpha_1\\ \alpha_2\\ ...\\\alpha_N).
\end{split}
\end{equation}
Denoting the smallest eigenvalue as $\lambda_0$, the ground state becomes $|GS\rangle = \frac{ \left( \alpha_0^{\dagger} \right)^N}{\sqrt{N!}}|0\rangle $. From the conditions for the particle number:
\begin{equation}
\begin{split}
	M_i
	&=\langle GS|R_i^{\dagger} R_i|GS \rangle =
	\langle 0|\frac{ \left( \alpha_0 \right)^N}{\sqrt{N!}} 
	\mqty(a^{\dagger} & R^{\dagger}_1& R^{\dagger}_2&...& R^{\dagger}_N)
	\mqty(\dmat{0,...,1,..., 0})	
	\mqty(a \\ R_1\\ R_2\\ ...\\R_N)
	\frac{ \left( \alpha_0^{\dagger} \right)^N}{\sqrt{N!}} |0\rangle \equiv
	\\
	&\equiv
	\langle 0|\frac{ \left( \alpha_0 \right)^N}{\sqrt{N!}} 
	\mqty(\alpha^{\dagger}_0 & \alpha^{\dagger}_1& ...& \alpha^{\dagger}_N)
	S\mqty(\dmat{0,...,1,..., 0})   S^\dagger 
	\mqty(\alpha_0\\ \alpha_1\\  ...\\\alpha_N)
	\frac{ \left( \alpha_0^{\dagger} \right)^N}{\sqrt{N!}}|0\rangle =
	\\
	&= \langle 0|\frac{ \left( \alpha_0 \right)^N}{\sqrt{N!}} 
	\left(\alpha^{\dagger}_0 \alpha_0   |S_{0,i}|^2 + \alpha^{\dagger}_1 \alpha_1 ...\right)
	\frac{ \left( \alpha_0^{\dagger} \right)^N}{\sqrt{N!}}|0\rangle
	= N |S_{0,i}|^2 =  N S_{0,i}^2,
\end{split}
\end{equation}
where the last equality follows from the Hamiltonian matrix being real and symmetric. Therefore $S_{0,i}^2 = M_i/N$. Similarly we get $S_{0,0}^2 = N_A/N$. It follows:
\begin{equation}
\begin{split}
	\langle H_I \rangle 
	&= \langle 0|\frac{ \left( \alpha_0 \right)^N}{\sqrt{N!}} \mqty(a^{\dagger} & R^{\dagger}_1& R^{\dagger}_2&...& R^{\dagger}_N) \mqty(0& g&g&...&g \\ g & 0&0&...&0\\g&0&0&...&0\\...&...\\g&0&0&...&0) \mqty(a \\ R_1\\ R_2\\ ...\\R_N) \frac{ \left( \alpha_0^{\dagger} \right)^N}{\sqrt{N!}} |0\rangle 
	\\
	&= 
	\langle 0|\frac{ \left( \alpha_0 \right)^N}{\sqrt{N!}}  \mqty(\alpha^{\dagger}_0 & \alpha^{\dagger}_1& \alpha^{\dagger}_2&...& \alpha^{\dagger}_N) S \mqty(0& g_1&g_2&...&g_M \\ g_1 & 0&0&...&0\\g_2&0&0&...&0\\...&...\\g_M&0&0&...&0)	 S^\dagger \mqty(\alpha_0\\ \alpha_1\\ \alpha_2\\ ...\\\alpha_N) \frac{ \left( \alpha_0^{\dagger} \right)^N}{\sqrt{N!}}|0\rangle
	\\
	&= \langle 0|\frac{ \left( \alpha_0 \right)^N}{\sqrt{N!}}  \mqty(\alpha^{\dagger}_0 & \alpha^{\dagger}_1& \alpha^{\dagger}_2&...& \alpha^{\dagger}_N) S
	\mqty(\sum_{1}^{M}g_i S_{0,i}& ...&...&...&... \\ S_{0,0}g_1 & ...&...&...&...\\S_{0,0} g_2&...&...&...&...\\...&...\\S_{0,0}g_M&...&...&...&...) \mqty(\alpha_0\\ \alpha_1\\ \alpha_2\\ ...\\\alpha_N) \frac{ \left( \alpha_0^{\dagger} \right)^N}{\sqrt{N!}}|0\rangle
	\\
	&=
	\langle 0|\frac{ \left( \alpha_0 \right)^N}{\sqrt{N!}} 
	\mqty(\alpha^{\dagger}_0 & \alpha^{\dagger}_1& \alpha^{\dagger}_2&...& \alpha^{\dagger}_N) \mqty(2S_{0,0}\sum_{1}^{M}g_i S_{0,i}& ...&...&...&... \\ ...& ...&...&...&...\\...&...&...&...&...\\...&...\\...&...&...&...&...)	 
	\mqty(\alpha_0\\ \alpha_1\\ \alpha_2\\ ...\\\alpha_N) \frac{ \left( \alpha_0^{\dagger} \right)^N}{\sqrt{N!}}|0\rangle
	\\
	&=
	2 N S_{0,0}\sum_{1}^{M}g_i S_{0,i}= 2 \sqrt{N_A}\sum_{1}^{M}g_i \sqrt{M_i}.
\end{split}
\end{equation}

\section{Appendix D: Two-Mode Hamiltonian and \texorpdfstring{$M$}{1} Harmonic Oscillators in the Environment}
\label{appendix:A3}
Similarly to the above calculation we can perform the calculation to get the ground state interaction energy in the case of the two-mode Hamiltonian in the phase accumulation stage with $N_A$ and $N_B$  particles in each well respectively, interacting with $M$ harmonic oscillators from the environment. The result is:
\begin{equation}
\langle H_I \rangle = 
2 (\sqrt{N_A}+\sqrt{N_B}) \sum_{i=1}^{M} g_i \sqrt{M_i}
\end{equation}
A direct way to obtain this is to do the same calculation as above. Another way is to observe that during the phase accumulation stage one has just two decoupled harmonic oscillators (diagonalizing $H_{TM}$), each interacting with the same bath, so the interaction energy is just the sum of those two contributions. This result can be generalized to a continuous bath of harmonic oscillators with the interaction Hamiltonian:
\begin{equation}
H_I=\int_0^{\infty} d\omega \rho(\omega) g(\omega) \left[\left(R^{\dagger}_\omega a +R_\omega^{\dagger} b\right) + h.c.\right],
\end{equation}
which results in the ground state interaction energy scaling as:
\begin{equation}
\langle H_I \rangle \propto
2 (\sqrt{N_A}+\sqrt{N_B}) \int_{0}^{\infty} d\omega \rho(\omega) g(\omega)  \sqrt{M_\omega}.
\label{res:appC}
\end{equation}
Another way to make sense of the above results is seeing them as a consequence of the coupling: raising/lowering operators. Namely, $a$ and $b$ contribute to the scaling as $\sqrt{N_A}$ and $\sqrt{N_B}$ respectively, and the operators $R_\omega^\dagger$ and $R_\omega$ contribute $\sqrt{M_\omega}$, where $N_A$, $N_B$ and $M_\omega$ are the occupation numbers of the left well, right well and the environment harmonic oscillator at frequency $\omega$, respectively.

If instead we have a different interaction Hamiltonian of the form:
\begin{equation}
H_I=\int_0^{\infty} d\omega \rho(\omega) g(\omega) \left(R^{\dagger}_\omega b^\dagger a +R_\omega a^\dagger b\right),
\end{equation}
and apply the same procedure, we arrive at the scaling of its ground state expectation value:
\begin{equation}
\langle H_I \rangle \propto
2 (\sqrt{N_A N_B}) \int_{0}^{\infty} d\omega \rho(\omega) g(\omega)  \sqrt{M_\omega}.
\label{res:apppC2}
\end{equation}

\section{Appendix E: Deriving the Lindblad Equation from the Microscopic Model}
\label{appendix:A4}
In this appendix, we derive the Lindblad master equation governing the reduced dynamics of the system using the microscopic approach detailed in~\cite{louisell1967solutions}.

{\it The Microscopic Model:} We consider the following total Hamiltonian:
\begin{equation}
H = H_S + H_E + H_I,
\end{equation}
where:
\begin{itemize}
\item $H_S$ is the system Hamiltonian;
\item $H_E$ describes the environment;
\item $H_I$ is the system-environment coupling Hamiltonian.
\end{itemize}

The system consists of two bosonic modes with annihilation operators $a$ and $b$, satisfying $[a, a^\dagger] = [b, b^\dagger] = 1$. The system Hamiltonian reads:
\begin{equation}
H_S = \frac{\delta}{2} \left( a^\dagger a - b^\dagger b \right) + V_0 \left( a^\dagger a + b^\dagger b \right),
\label{System_hamiltonian}
\end{equation}
where $\delta$ is the energy difference between the modes and $-V_0$ is the chemical potential. The environment is modeled as a continuum of bosonic modes with annihilation operators $R_\omega$ and density of states $\rho(\omega)$. Its Hamiltonian is
\begin{equation}
H_E = \int_0^\infty d\omega \, \rho(\omega) \, \omega \, R_\omega^\dagger R_\omega.
\end{equation}
The considered coupling Hamiltonian is
\begin{equation}
H_I = \int_0^\infty d\omega \, \rho(\omega) \, g(\omega) \left( R_\omega^\dagger (a + b) + (a^\dagger + b^\dagger) R_\omega \right),
\end{equation}
where $g(\omega)$ is the coupling strength to the environment.

{\it Interaction Picture and Operator Decomposition:} We work in the interaction picture with respect to $H_0 = H_S + H_E$. The interaction picture operators are defined as:
\begin{equation}
O_I(t) = e^{i H_0 t} O e^{-i H_0 t}.
\end{equation}
First, we note the following commutation relations:
\begin{equation}
[H_S, a] = - \left( \frac{\delta}{2} + V_0 \right) a, \quad [H_S, b] = \left( \frac{\delta}{2} - V_0 \right) b.
\end{equation}
Thus, the time evolution of $a$ and $b$ in the interaction picture is
\begin{align}
a(t) &= a \, e^{-i \left( V_0 + \frac{\delta}{2} \right) t}, \\
b(t) &= b \, e^{-i \left( V_0 - \frac{\delta}{2} \right) t}.
\end{align}
Similarly, the bath operators evolve as:
\begin{equation}
R(\omega)(t) = R_\omega \, e^{-i \omega t}.
\end{equation}
The interaction Hamiltonian in the interaction picture becomes:
\begin{equation}
H_I(t) = \int_0^\infty d\omega \, \rho(\omega) \, g(\omega) \left[ R_\omega^\dagger(t) \left( a(t) + b(t) \right) + \mathrm{h.c.} \right].
\end{equation}
Substituting the explicit time evolution we have:
\begin{align}
H_I(t) = \int_0^\infty d\omega \, \rho(\omega) \, g(\omega) \Big[ & R_\omega^\dagger e^{i \omega t} \left( a \, e^{-i \left( V_0 + \frac{\delta}{2} \right) t} + b \, e^{-i \left( V_0 - \frac{\delta}{2} \right) t} \right) \nonumber \\
& + \left( a^\dagger \, e^{i \left( V_0 + \frac{\delta}{2} \right) t} + b^\dagger \, e^{i \left( V_0 - \frac{\delta}{2} \right) t} \right) R_\omega \, e^{-i \omega t} \Big].
\end{align}
Expanding and collecting terms:
\begin{align}
H_I(t) = \int_0^\infty d\omega \, \rho(\omega) \, g(\omega) \Big[ & R_\omega^\dagger \, e^{i (\omega - (V_0 + \delta/2)) t} \, a + R_\omega^\dagger \, e^{i (\omega - (V_0 - \delta/2)) t} \, b \nonumber \\
& + a^\dagger \, e^{i (-\omega + V_0 + \delta/2) t} \, R_\omega + b^\dagger \, e^{i (-\omega + V_0 - \delta/2) t} \, R_\omega \Big].
\end{align}

{\it The Master Equation:} Using the Born-Markov approximation \cite{breuer2002theory}, the reduced system dynamics is given by:
\begin{equation}
\frac{d}{dt} \rho(t) = - \int_0^\infty ds \, \mathrm{Tr}_E \left[ H_I(t), \left[ H_I(t - s), \rho(t) \otimes \rho_E \right] \right],
\end{equation}
where $\rho_E$ is the equilibrium state of the environment. We evaluate this expression by plugging in the $H_I(t)$ and using standard correlation functions for the bosonic bath:
\begin{align}
& \mathrm{Tr}_E\left( R_\omega^\dagger(t) R_{\omega'}(t - s) \rho_E\right) =\langle R_\omega^\dagger(t) R_{\omega'}(t - s) \rangle =\langle R_\omega^\dagger(s) R_{\omega'} \rangle, \\
& \langle R_\omega^\dagger(s) R_{\omega'} \rangle=  \delta(\omega - \omega') \, n(\omega) \, e^{i \omega s}/\rho(\omega) \label{eq:bath_correlation1}, \\
&\langle R_\omega(s) R_{\omega'}^\dagger \rangle = \delta(\omega - \omega') \, \left( n(\omega) + 1 \right) e^{-i \omega s}/\rho(\omega), 
\label{eq:bath_correlation2}
\end{align}
where $n(\omega)$ is the Bose-Einstein distribution
\begin{equation}
n(\omega) = \frac{1}{e^{\beta \omega} - 1}
\end{equation}
and $\beta = 1/(k_B T)$. After evaluating the commutators and performing the $s$-integral, the master equation takes the standard Lindblad form:
\begin{equation}
\frac{d}{dt} \rho = -i \left[ H_S + H_{\mathrm{LS}}, \rho \right] + \mathcal{D}[\rho],
\end{equation}
where $H_{\mathrm{LS}}$ is the Lamb shift Hamiltonian and the dissipator reads:
\begin{align}
\mathcal{D}[\rho] &= \gamma_a \left( n_a + 1 \right) \left( a \rho a^\dagger - \frac{1}{2} \left\{ a^\dagger a, \rho \right\} \right) + \gamma_a n_a \left( a^\dagger \rho a - \frac{1}{2} \left\{ a a^\dagger, \rho \right\} \right) \nonumber \\
&+ \gamma_b \left( n_b + 1 \right) \left( b \rho b^\dagger - \frac{1}{2} \left\{ b^\dagger b, \rho \right\} \right) + \gamma_b n_b \left( b^\dagger \rho b - \frac{1}{2} \left\{ b b^\dagger, \rho \right\} \right).
\end{align}
Here:
\begin{itemize}
\item $\omega_a = V_0 + \frac{\delta}{2}$ and $\omega_b = V_0 - \frac{\delta}{2}$ are the effective transition frequencies.
\item The decay rates are:
\begin{equation}
	\gamma_a = 2\pi \rho(\omega_a) |g(\omega_a)|^2, \quad \gamma_b = 2\pi \rho(\omega_b) |g(\omega_b)|^2.
\end{equation}
\item The thermal occupation numbers are:
\begin{equation}
	n_a = \frac{1}{e^{\beta \omega_a} - 1}, \quad n_b = \frac{1}{e^{\beta \omega_b} - 1}.
\end{equation}
\end{itemize}

{\it High-Frequency Approximation:} 
We now consider the physically relevant regime where both the detuning $\delta$ and the environment thermal energy scale $k_B T$ are much smaller than the chemical potential scale $V_0$, i.e.,
\begin{equation}
\delta \ll V_0, \quad k_B T \ll V_0.
\end{equation}
In this limit, the two system transition frequencies become approximately equal:
\begin{equation}
\omega_a = V_0 + \frac{\delta}{2} \approx V_0, \quad \omega_b = V_0 - \frac{\delta}{2} \approx V_0.
\end{equation}
Consequently, the decay rates simplify as
\begin{equation}
\gamma_a \approx \gamma_b\approx 2 \pi \rho(V_0) |g(V_0)|^2,
\end{equation}
and the thermal occupation numbers become negligible:
\begin{equation}
n_a \approx n_b \approx \frac{1}{e^{\beta V_0} - 1} \ll 1.
\end{equation}
Thus, by using $\alpha \equiv (a+b)/\sqrt 2$ and $\gamma = 4 \pi \rho(V_0) |g(V_0)|^2$, the dissipator reduces to:
\begin{equation}
\mathcal{D}[\rho] = \gamma \left( \alpha \rho \alpha^\dagger - \frac{1}{2} \left\{ \alpha^\dagger \alpha, \rho \right\} \right)
\end{equation}

Physically, this approximation reflects that the environment cannot resolve the small energy difference between the two modes, leading to nearly identical dissipative dynamics for both.

{\it A Number-conserving Interaction:}
We now consider a modified interaction Hamiltonian of the form
\begin{equation}
H_I = \int_0^\infty d\omega \, \rho(\omega) \, g(\omega) \left( R_\omega a^\dagger b + R_\omega^\dagger b^\dagger a \right).
\end{equation}
To obtain the Lindblad operators, we here follow the formalism of \cite{breuer2002theory}, which employs the Born-Markov approximation. Defining the system and the environment operators as
\begin{equation}
J_+ \equiv a^\dagger b
, \quad J_- \equiv b^\dagger a
,\quad G^\dagger  \equiv \int_0^\infty d\omega \, \rho(\omega) \, g(\omega)  R_\omega^\dagger 
,\quad G  \equiv \int_0^\infty d\omega \, \rho(\omega) \, g(\omega)  R_\omega
\end{equation}
we can rewrite the interaction Hamiltonian as
\begin{equation}
H_I = J_+ G + J_- G^\dagger = (J_++J_-)\frac{G+G^\dagger}{2}+(i(J_+-J_-))\frac{i(G^\dagger-G)}{2}\equiv A_1 B_1 + A_2B_2 ,
\end{equation}
where we have constructed hermitian operators $A_1$, $A_2$, $B_1$ and $B_2$ . Passing to the interaction picture with respect to $H_0$, we first note that:
\begin{equation}
J_+ (t)= J_+ e^{i\delta t}
, \quad J_- (t)= J_- e^{-i\delta t}
,\quad G^\dagger(t) =\int_0^\infty d\omega \, \rho(\omega) \, g(\omega)  R_\omega^\dagger e^{i \omega t} 
,\quad G(t) =\int_0^\infty d\omega \, \rho(\omega) \, g(\omega)  R_\omega e^{-i \omega t} ,
\end{equation}
where we used the commutation relation with (\ref{System_hamiltonian}). The operator \(a^\dagger a + b^\dagger b\) commutes with \(J_\pm\) and does not contribute to its time evolution. Using (\ref{eq:bath_correlation1}) and (\ref{eq:bath_correlation2}), we can evaluate: 
\begin{equation}
\langle G^\dagger(s) G \rangle = \int_0^\infty d\omega \rho(\omega) g(\omega)^2  n(\omega) e^{i\omega s} 
, \quad
\langle G(s) G^\dagger \rangle = \int_0^\infty d\omega \rho(\omega) g(\omega)^2 (1+ n(\omega)) e^{-i\omega s} .
\label{eq:correlation_G}
\end{equation}
We proceed by constructing the operators $\gamma_{\alpha\beta}(\tilde \omega)$  as:
\begin{equation}
\gamma_{\alpha\beta} (\tilde \omega) \equiv \int_{-\infty}^\infty ds e^{i \tilde \omega s } \langle B_\alpha (s) B_\beta \rangle .
\end{equation}
Plugging in the expressions for $B_\alpha$,  (\ref{eq:correlation_G}), and noting that $n(\tilde \omega)=0$ for $\tilde \omega <0$, we get:
\begin{equation}
\begin{split}
& \gamma_{11} (\tilde \omega) = \int_{-\infty}^\infty ds e^{i \tilde \omega s } \langle B_1 (s) B_1 \rangle  = \int_{-\infty}^\infty ds e^{i \tilde \omega s } \frac{1}{4} ( \langle G (s) G^\dagger \rangle + \langle G ^\dagger(s) G \rangle) 
=\frac{\pi}{2} \rho(\tilde \omega) g( \tilde\omega )^2 (n(-\tilde \omega) + (1+n(\tilde \omega))) =\gamma_{22} (\tilde \omega) \\
&\gamma_{12}(\tilde \omega) = \frac{i \pi}{2} \rho(\tilde\omega) g(\tilde\omega)^2 ((1+n(\tilde\omega)-n(-\tilde\omega)) = -\gamma_{21}(\tilde \omega). \nonumber
\end{split}
\end{equation}
As in \cite{breuer2002theory}, we first decompose the operators $A_\alpha = \sum_\omega A_\alpha (\omega)$ where $A_\alpha (\omega)$  is defined by the condition $[H_S, A_\alpha(\omega)]=-\omega A_\alpha(\omega)$. One can easily see that:
\begin{equation}
\begin{split}
    & A_1 =A_1(\delta)+A_1(-\delta) = J_- + J_+  ,\\
    & A_2 =A_2(\delta)+A_2(-\delta) = -iJ_- +iJ_+ .
\end{split}
\end{equation}
Finally we can evaluate the dissipator in the Schrödinger picture as:
\begin{equation}
    \mathcal{D}(\rho) = \sum_{\omega} \sum_{\alpha,\beta} \gamma_{\alpha\beta}(\omega)
    \left(
        A_{\beta}(\omega)\,\rho A_{\alpha}^{\dagger}(\omega)
        - \frac{1}{2} \left\{ A_{\alpha}^{\dagger}(\omega) A_{\beta}(\omega), \rho \right\}
    \right),
    \label{eq:placeholder_label2}
\end{equation}
where $\omega \in \{ -\delta, \delta\}$ and $\alpha, \beta \in \{ 1, 2\}$. Thus expanding the summation:
\begin{equation}
\begin{split}
\mathcal{D}[\rho] =  &\, \gamma_{11}(\delta)\left( A_1(\delta) \rho A_1^\dagger(\delta)-  \frac{1}{2} \{ A_1^\dagger(\delta) A_1(\delta), \rho \} \right) \nonumber + \, \gamma_{22}(\delta)\left( A_2(\delta) \rho A_2^\dagger(\delta)-  \frac{1}{2} \{ A_2^\dagger(\delta) A_2(\delta), \rho \} \right) \\
 + & \, \gamma_{12}(\delta)\left( A_2(\delta) \rho A_1^\dagger(\delta)-  \frac{1}{2} \{ A_1^\dagger(\delta) A_2(\delta), \rho \} \right) + \, \gamma_{21}(\delta)\left( A_1(\delta) \rho A_2^\dagger(\delta)-  \frac{1}{2} \{ A_2^\dagger(\delta) A_1(\delta), \rho \} \right) \\
 +&\, \gamma_{11}(-\delta)\left( A_1(-\delta) \rho A_1^\dagger(-\delta)-  \frac{1}{2} \{ A_1^\dagger(-\delta) A_1(-\delta), \rho \} \right) + \, \gamma_{22}(-\delta)\left( A_2(-\delta) \rho A_2^\dagger(-\delta)-  \frac{1}{2} \{ A_2^\dagger(-\delta) A_2(-\delta), \rho \} \right) \\
 + & \, \gamma_{12}(-\delta)\left( A_2(-\delta) \rho A_1^\dagger(-\delta)-  \frac{1}{2} \{ A_1^\dagger(-\delta) A_2(-\delta), \rho \} \right) + \, \gamma_{21}(-\delta)\left( A_1(-\delta) \rho A_2^\dagger(-\delta)-  \frac{1}{2} \{ A_2^\dagger(-\delta) A_1(-\delta), \rho \} \right),
 \end{split}
\end{equation}
and plugging everything in, we obtain the dissipator in the Lindblad form:
\begin{equation}
\mathcal{D}[\rho] = \, \gamma \, n(\delta)\left( J_+ \rho J_- - \frac{1}{2} \{ J_- J_+, \rho \} \right) + \gamma \,(n(\delta)+1) \left( J_- \rho J_+ - \frac{1}{2} \{ J_+ J_-, \rho \} \right),
    \label{eq:conserving_dissipator}
\end{equation}
where we introduced the dissipation rate:  
\[
\gamma = 2\pi \rho(\delta) |g(\delta)|^2.
\]
The Lamb shift Hamiltonian \(H_{\mathrm{LS}}\) arises similarly from the principal value integrals but is omitted here for brevity. This number-conserving interaction describes excitation exchange between the modes \(a\) and \(b\) mediated by the environment. Under the low-temperature assumption \(k_B T \ll \delta\), the thermal occupation number:
\[
n(\delta) = \frac{1}{e^{\beta \delta} - 1} \ll 1,
\]
and we can neglect terms proportional to \(n(\delta)\), yielding
\begin{equation}
\mathcal{D}[\rho] \approx \gamma \left( J_- \rho J_+ - \frac{1}{2} \{ J_+ J_-, \rho \} \right).
\end{equation}
This effectively describes zero-temperature damping with only spontaneous emission-like processes.

\end{document}